\documentclass[preprint]{aastex}

\usepackage{rotating}
\usepackage{subfigure}
\usepackage{grffile} 
\usepackage{url}
\usepackage{standalone}
\standalonetrue
\usepackage{import}
\usepackage[utf8]{inputenc}
\usepackage{booktabs}

\usepackage{color}
\usepackage{xspace}
\usepackage[usenames,dvipsnames]{xcolor}

\newcommand{\okinfinal}[1]{\textcolor{red}{#1}}

\newcommand{\hh}{\ensuremath{\textrm{H}_{2}}}			
\newcommand{\percc}{\ensuremath{\textrm{cm}^{-3}}}

\newcommand{\um}{\ensuremath{\mu \textrm{m}}\xspace}    

\newcommand{\twelveco}{\ensuremath{^{12}\textrm{CO}}\xspace}
\newcommand{\thirteenco}{\ensuremath{^{13}\textrm{CO}}\xspace}

\newcommand{\degrees}{\ensuremath{^{\circ}}}

\newcommand{\vone}{{\rm v}1.0\xspace}
\newcommand{\vtwo}{{\rm v}2.0\xspace}
\newcommand\mjysr{\ensuremath{{\rm MJy~sr}^{-1}}}

\newcommand\nbolocat{8594\xspace}
\newcommand\nbolocatnew{591\xspace}
\renewcommand\arcdeg{\mbox{$^\circ$}\xspace} 
\renewcommand\arcmin{\mbox{$^\prime$}\xspace} 
\renewcommand\arcsec{\mbox{$^{\prime\prime}$}\xspace} 
\def\secref#1{Section \ref{#1}}

\def\Figure#1#2#3#4#5{
\begin{figure*}[htp]
\includegraphics[scale=#4,angle=#5]{#1}
\caption{#2}
\label{#3}
\end{figure*}
}

\def\FigureTwo#1#2#3#4#5{
\begin{figure*}[htp]
\epsscale{#5}
\plottwo{#1}{#2}
\caption{#3}
\label{#4}
\end{figure*}
}

\def\TallFigureTwo#1#2#3#4#5#6{
\begin{figure*}[htp]
\epsscale{#5}
\subfigure[]{ \includegraphics[width=#6]{#1} }
\subfigure[]{ \includegraphics[width=#6]{#2} }
\caption{#3}
\label{#4}
\end{figure*}
}

\def\Table#1#2#3#4#5#6{
\begin{deluxetable}{#1}
\tablewidth{0pt}
\tabletypesize{\footnotesize}
\tablecaption{#2}
\tablehead{#3}
\startdata
\label{#4}
#5
\enddata
\bigskip
#6
\end{deluxetable}
}

\shorttitle{BGPS VIII: Outer Galaxy}
\shortauthors{Ginsburg et al}

\begin{document}

\standalonefalse
\renewcommand{\okinfinal}[1]{{#1}}

\title{The Bolocam Galactic Plane Survey IX: Data Release 2 and Outer Galaxy Extension}

\newcommand{\casa}{1}
\newcommand{\uab}{2}
\newcommand{\yale}{3}
\newcommand{\utexas}{4}
\newcommand{\arizona}{5}
\newcommand{\penn}{6}

\author{Adam Ginsburg\altaffilmark{\casa},
        Jason Glenn\altaffilmark{\casa},
        Erik Rosolowsky\altaffilmark{\uab},
        Timothy P. Ellsworth-Bowers\altaffilmark{\casa},
        Cara Battersby\altaffilmark{\casa},
        Miranda Dunham\altaffilmark{\yale},
        Manuel Merello\altaffilmark{\utexas},
        Yancy Shirley\altaffilmark{\arizona},
        John Bally\altaffilmark{\casa},
        Neal J. Evans II\altaffilmark{\utexas},
        Guy Stringfellow\altaffilmark{\casa},
        James Aguirre\altaffilmark{\penn}}
\email{Adam.Ginsburg@colorado.edu}

\affil{{$^\casa$}{\it{CASA, University of Colorado, 389-UCB, Boulder, CO 80309}}}

\affil{{$^{\uab}$}{\it{ 
Department of Physics
4-181 CCIS
University of Alberta
Edmonton AB T6G 2E1
CANADA
}}}

\affil{{$^{\yale}$}{\it{ Department of Astronomy, Yale University
P.O. Box 208101, New Haven, CT, 06520}}}

\affil{{$^\utexas$}{\it{
The University of Texas
Department of Astronomy
2515 Speedway, Stop C1400
Austin, Texas 78712-1205
}}}

\affil{{$^{\arizona}$}{\it{ Steward Observatory, University of
Arizona, 933 North Cherry Avenue, Tucson, AZ 85721 }}}

\affil{{$^{\penn}$}{\it{ Department of Physics and Astronomy, University of
Pennsylvania, 209 South 33rd Street, Philadelphia, PA 19104 }}}


\begin{abstract}

We present a re-reduction and expansion of the Bolocam Galactic Plane Survey,
first presented by \citet{Aguirre2011} and \citet{Rosolowsky2010}.  The BGPS is
a 1.1 mm survey of dust emission in the Northern galactic plane, covering
longitudes $-10\degrees < \ell < 90\degrees$ and latitudes $|b| < 0.5\degrees$
with a typical $1-\sigma$ RMS sensitivity of 30-100 mJy in a $\sim33\arcsec$
beam.  Version 2 of the survey includes an additional $\sim20$ square degrees
of coverage in the 3rd and 4th quadrants and $\sim2$ square degrees in the 1st
quadrant.  The new data release has improved angular recovery, with complete
recovery out to $\sim80\arcsec$ and partial recovery to $\sim300\arcsec$, and
reduced negative bowls around bright sources resulting from the atmospheric
subtraction process.  We resolve the factor of 1.5 flux calibration offset
between the \vone\ data release and other data sets and determine that there is
no offset between \vtwo and other data sets.  The \vtwo pointing accuracy is
tested against other surveys and demonstrated to be accurate and an improvement
over \vone.  We present simulations and tests of the pipeline and its
properties, including measurements of the pipeline's angular transfer function.  

The Bolocat cataloging tool was used to extract a new catalog, which includes
\nbolocat\ sources, with \nbolocatnew in the expanded regions.  We have
demonstrated that the Bolocat 40\arcsec and 80\arcsec apertures are accurate
even in the presence of strong extended background emission.  The number of
sources is lower than in \vone, but the amount of flux and area included in
identified sources is larger.

\end{abstract}

\section{Introduction}

Observations in the millimeter continuum provide the best method to identify
a Galaxy-wide sample of star-forming clumps.  The emission is optically
thin, minimally affected by temperature and can be surveyed over large areas.
Unlike targeted observations, blind surveys allow for a complete and
systematic study of dense gas clumps.

In the past decade, there have been many Galactic plane surveys at
millimeter/submillimeter wavelengths, of which the Bolocam Galactic Plane
Survey \citep{Aguirre2011,Rosolowsky2010a} was the first to be completed and
publicly released.  ATLASGAL \citep{Schuller2009a} surveyed the southern
Galactic plane at 870 \um.  The JCMT Galactic Plane Survey, or JPS, will survey
the Northern plane at 850 \um.  In the past 3 years, the Hi-Gal Galactic Plane
survey observed the Galaxy from 70 to 500 \um with the Herschel Space
Observatory, sensing the peak of the dust SED with minimal spatial filtering
\citep{Molinari2010}.  The survey has provided access to the peak of the dust
spectral energy distribution at modest ($\lesssim40\arcsec$) resolution
\citep{Traficante2011a}.  Together, these surveys provide a complete map of
long-wavelength dust emission across the Galactic plane.

Long-wavelength data
are essential for constraining the dust emissivity, one of the free parameters
in greybody spectral energy distribution (SED) fits.
\citet{Shetty2009b,Shetty2009a} demonstrated the need for long-wavelength data
to accurately determine both $\beta$, the dust emissivity spectral index, and
temperature.  \citet{Juvela2012d} also showed that adding additional
wavelengths to an SED fit, even with lower signal-to-noise, significantly
reduces the degeneracy in the fit.

Millimeter-wave dust emission also has the advantage of being relatively
insensitive to temperature.  When looking at cold gas, $T\lesssim20$ K, all of
the Herschel bands deviate from a Rayleigh-Jeans temperature approximation.
Longer wavelength observations are less affected by temperature assumptions.
The  1.1 mm band is in many cases the longest wavelength unaffected by
free-free emission, providing the least environmentally-biased view of
optically thin dust emission and therefore total dust mass.

Similarly, at millimeter wavelengths, the dust opacity is low enough that all
clumps detected in the BGPS are expected to be optically thin \citep[with the
possible exception of Sgr B2;][]{Bally2010a}.  In combination with the weak
temperature dependence, this feature of 1.1 millimeter emission allows for the
most direct and straightforward estimates of total dust mass. 

Millimeter-bright dust clumps are generally associated with high-density,
star-forming gas.  Previous surveys have found cold, massive molecular clouds
via the \twelveco and \thirteenco 1-0 lines \citep{Dame2001a,Jackson2006a}.
However, these clouds are only moderate density, $n(\hh)\sim10^2-10^3$ \percc, while
dust-detected clumps have typical densities $n(\hh)\gtrsim10^4$ \percc
\citep{Dunham2010a}.  The dense gas in these clumps is more directly associated
with star and cluster formation \citep{Dunham2011b, Battersby2010a}, allowing
for systematic studies of pre-star-forming and star-forming gas.

The BGPS \vone data has been public since 2009, and has been used extensively
as both a finder chart and a tool to probe Galactic properites.  It was used to
examine the properties of maser sources \citep{Pandian2012,Chen2012}, outflow
sources \citep{Ioannidis2012}, and high-mass star-forming regions
\citep{Reiter2011,Battersby2011,Dunham2011}.  It has served as the basis for
studies of forming clusters \citep{Alexander2012a,Ginsburg2012a} and
intermediate-mass stars \citep{Arvidsson2010a}.  The BGPS and other surveys
have served as finder charts for large-scale millimeter line studies of the
Galactic plane \citep{Schenck2011,Schlingman2011,Ginsburg2011a,Shirley2013a}.
BGPS clumps have been used as the target sample for distance determinations to
large cloud populations \citep{Ellsworth-Bowers2013a}.  \citet{Dunham2011b}
used the BGPS to measure properties of star forming regions as a function of
Galactocentric radius.  These and many other ongoing and planned studies
demonstrate the need for, and benefits of, publicly available blind legacy
surveys.

This paper presents \vtwo of the Bolocam Galactic Plane Survey (BGPS), with a
complete data release available at
\url{irsa.ipac.caltech.edu/data/BOLOCAM_GPS/}.  In Paper I \citep{Aguirre2011},
the initial processing of the BGPS \vone was described in detail.  It was noted
in Section 5 of \citet{Aguirre2011} that there was a discrepancy between our
survey and previously published results.  This discrepancy raised the
possibility of a flux calibration error in the Version 1 (hereafter \vone)
results: we confirm and correct the error in this paper.  In addition, we have
made significant improvements to the data pipeline, measured important features
of the pipeline including its angular transfer function, improved the pointing
accuracy, and added new observations.

The paper is as follows:
We resolve the flux calibration discrepancy in \okinfinal{\secref{sec:calibration}.
In \secref{sec:observations}, we discuss new observations included in the \vtwo data.
\secref{sec:datareduction} describes changes to the data reduction process and new
data products.  \secref{sec:stf} and \secref{sec:sourceextraction} measure the
angular transfer function of the BGPS \vtwo pipeline and properties of
extracted sources respectively.}   The paper concludes with a brief discussion
of the results and a summary.

\section{Calibration}
\label{sec:calibration}
The original calibration, along with tables of color correction and a detailed
treatment of the filter response, are described in \citet{Aguirre2011a}.
We discuss important changes in \vtwo in this section.

\subsection{Why was there a multiplicative offset in the \vone data release?}
\label{sec:caloffset}
In \citet{Aguirre2011}, we reported that a `correction factor' of about 1.5 on average was
needed to bring our data into agreement with other 1 mm data sets.
We discovered that the published \vone BGPS images have
a different calibration reported in their FITS headers than was used in
processing the data.  The calibration used in the released data was 
borrowed from a previous observing run, during which a different bias voltage
was used, and differed from the pipeline-derived calibration by a factor
$\approx 1.5$, completely explaining the discrepancy.

\subsection{Comparing \vone and \vtwo calibration}
\label{sec:v1v2compare}
We checked the data for consistency with the measured calibration offset.  In
order to compare flux densities in identical sources, we performed aperture
photometry on the \vtwo data based on the locations of \vone sources using both
the `source masks' from Bolocat \vone \citep{Rosolowsky2010} and circular
apertures centered on the Bolocat \vone peaks.  Source masks, also known as label masks,
are images in which the value of a pixel is either 0 for no source or the catalog number
if there is a source associated with that pixel.

\mathchardef\mhyphen="2D

We measured the
multiplicative offset between \vone\ and \vtwo by comparing these aperture-extracted fluxes. 
For each aperture size, we 
measured the best-fit line between the \vone and \vtwo data
using a total least squares
(TLS%
\footnote{\url{https://code.google.com/p/agpy/source/browse/trunk/agpy/fit_a_line.py}%
, see also
\url{http://astroml.github.com/book_figures/chapter8/fig_total_least_squares.html}})
method weighted by the flux measurement errors as reported in the catalogs.
The agreement with $S_{\vtwo}=1.5~S_{\vone}$, as expected based on \okinfinal{Section \ref{sec:caloffset}},
is within 10\%, although  
the larger apertures show a slight excess with
$S_{\vtwo}\approx(1.6\mhyphen1.7)~S_{\vone}$.  This excess is expected given the improved
extended flux recovery in \vtwo (see Section \ref{sec:stf}).  The \vtwo/\vone\ flux ratio is weakly
dependent on the source flux density, with higher \vtwo/\vone\ ratios for
brighter sources.

\subsection{Comparison to Other Surveys}
In Section 5.5 of \citet{Aguirre2011}, we compared the BGPS \vone\ data to other
data sets from similar-wavelength observations.  We repeat those comparisons
here using the \vtwo\ data and demonstrate that \vtwo achieves better agreement
with other data sets than \vone.  Full details of the comparison were given in
\citet{Aguirre2011a}.

We compare to 3 data sets in the same $\sim 1$ mm atmospheric window.  Two data
sets from MAMBO II, the \citet[][M07]{Motte2007} Cygnus X survey and the
\citet[][R06]{Rathborne2006} IRDC survey, overlap with the BGPS.  The SIMBA 1.3
mm survey of the $\ell=44\degrees$ region is the largest survey in the 1 mm
band that overlaps with ours \citep[][M09]{Matthews2009}.

The comparison data sets have angular transfer functions that differ from the
BGPS.  In order to account for the difference, we allow for a large angular
scale offset between the observations.  We fit a line of the form $y=m x + b$
to the data, where $x$ and $y$ represent the pixel values gridded to 7.2\arcsec
pixels.  The $b$ value allows for a local offset, i.e. a non-zero $b$ value
indicates a substantial difference in the angular transfer function.  Since
such an additive offset is unlikely to apply across the entire observed region,
we also fit the offset for small sub-regions in the M07 and M09 data, focusing
on DR21 and a region centered on G45.5+0.1 respectively.

\Table
{lrrrrrr}
{Flux comparison with R06, M07, and M09}
{Comparison & \multicolumn{2}{c}{Pixels      } & \multicolumn{2}{c}{Pixels       }   & \multicolumn{2}{c}{Pixels       }  \\
Survey      & \multicolumn{2}{c}{$>3~\mjysr$ } & \multicolumn{2}{c}{$>10~\mjysr$ }   & \multicolumn{2}{c}{$>20~\mjysr$ }  \\ 
\hline
BGPS \vone  & \colhead{$m$}   &\colhead{$b$}& \colhead{$m$}   & \colhead{$b$} & \colhead{$m$}   & \colhead{$b$} \\
\cmidrule(l){2-3}
\cmidrule(l){4-5}
\cmidrule(l){6-7}
}
{tab:FluxComparison}
{
R06     & 1.39 & $-2.00$ & 1.46 & $-2.79$ & 1.53 & $-4.77$ \\  
M07     & 1.51 & $4.13 $ & 1.44 & $13.78$ & 1.36 & $27.45$ \\ 
M07DR21 & 1.36 & $28.03$ & 1.31 & $37.91$ & 1.25 & $49.44$ \\
M09     & 1.32 & $-0.22$ & 1.25 & $4.94 $ & 1.21 & $ 9.88$ \\
M09a    & 1.50 & $-5.15$ & 1.51 & $-4.82$ & 1.53 & $-5.11$ \\
\hline 
\hline \\
\multicolumn{7}{l}{BGPS v2} \\
\hline \\
R06     & 1.05 & $3.67 $ & 1.02 & $5.03 $ & 1.00 &  7.05 \\  
M07     & 1.16 & $6.51 $ & 1.12 & $12.75$ & 1.08 & 21.04 \\
M07DR21 & 1.09 & $21.98$ & 1.07 & $27.61$ & 1.04 & 34.21 \\
M09     & 0.73 & $1.33 $ & 0.69 & $6.75 $ & 0.66 & 13.45 \\
M09a    & 0.96 & $-3.21$ & 0.94 & $-0.69$ & 0.89 & 2.91  \\
}{\\
The table values $m$ and $b$ are given for a linear fit of the form $y=mx+b$. \\
References: \citet[][R06; MAMBO]{Rathborne2006},
\citet[][M07; MAMBO]{Motte2007},
\citet[][M09; SIMBA]{Matthews2009}.\\
M09a refers to the G45.5+0.1 region,
and M07DR21 refers to the DR21 region
}

The results of that comparison are displayed in Table \ref{tab:FluxComparison}, 
which includes the original comparison from \citet{Aguirre2011}.%
\footnote{In \citet{Aguirre2011}, there was a minor error in the table: M07 and
M09 were swapped.  This has been corrected in Table \ref{tab:FluxComparison}.}
BGPS \vtwo is in much better agreement with the other data sets than \vone, but
it retains a significant additive offset, particularly with respect to MAMBO.
The additive offset is explained by a difference in the angular transfer
function; the MAMBO observing strategy of fast position switching allows
structures on the scale of the array to be preserved, while Bolocam's fast-scan
strategy does not.  The differing observing strategy explains why there is an
additive offset between Bolocam and MAMBO, but no such offset for SIMBA, which
was used in a fast-scan mode similar to Bolocam.  The varying backgrounds in
separate regions account for some of the remaining multiplicative offset.  When
individual sub-regions are compared, the additive and multiplicative offsets
more clearly separate into independent components, i.e. a line with an additive
offset is a better fit to the data.

To enable a comparison of flux density between the surveys, we must account for
the different spectral bandpasses of the instruments.  The relative flux
density measured between the instruments depends on the spectral index
$\alpha_{\nu}$ of the observed source; $\alpha_{\nu}=2$ corresponds to a
perfect black body on the Rayleigh-Jeans tail.%
\footnote{If the underyling spectral indices of the emission regions are
uncorrelated with the flux, e.g.  if they are constant, the slopes in Table
\ref{tab:FluxComparison} will be unaffected.  The assumption of constant
spectral index with flux is reasonable since observed spectral index-flux
correlations are shallow \citep{Kelly2012a}.  }
In Table \ref{tab:relflux} we show the relative flux densities expected for
Bolocam, MAMBO and SIMBA; they differ by at most 19\% for spectral indices
$\alpha_{\nu}<5$.  Bolocam flux densities are expected to be higher because
Bolocam has a higher effective central frequency than either of the other
instruments.

In \citet{Aguirre2011}, we measured Bolocam \vone/MAMBO and Bolocam \vone/SIMBA ratios in
the range $0.66 < R < 0.83$, indicating a clear disagreement between the
surveys.  With the \vtwo\  data, we measure ratios $0.97 < R_{SIMBA} < 1.08$ and
$0.89 < R_{MAMBO} < 0.99$.  These numbers still indicate that the BGPS is too
faint by $\lesssim20\%$ relative to the expectations laid out in Table
\ref{tab:relflux}, but with a systematic calibration error no better than 20\%
in each survey, this level of agreement is reasonable.

\Table{lccccc}
{Relative flux measurements of Bolocam, MAMBO, and SIMBA for different input sources}
{  $\alpha$ &    Bolocam/MAMBO &   Bolocam/SIMBA
\\ }
{tab:relflux}{
       1.0 &            1.092 &           1.096 \\
       1.5 &            1.092 &           1.096 \\
       2.0 &            1.092 &           1.096 \\
       2.5 &            1.110 &           1.119 \\
       3.0 &            1.126 &           1.140 \\
       3.5 &            1.138 &           1.157 \\
       4.0 &            1.146 &           1.170 \\
       4.5 &            1.151 &           1.181 \\
       5.0 &            1.152 &           1.188 \\
}{\\
Response functions are computed using an atmospheric transmittance
of 1 mm of precipitable water vapor.}

\section{Expansion of the BGPS and Observations}
\label{sec:observations}
Thirteen nights of additional data were acquired from December 15th, 2009 to
January 1st, 2010.  The target fields and areas covered are listed in Table
\ref{tab:observations} as boxes in Galactic latitude and longitude, with
position angles to the Galactic plane indicated.  The original
observations are described in Section 2 of \citet{Aguirre2011}.

\Table{lccccc}
{Observations}
{Target & Longitude & Latitude & Longitude Size & Latitude Size & Position Angle}
{tab:observations}
{
               IRAS 22172 &     102.91  &      -0.64  &       1.67  &       1.07  &      0 \\
                     l106 &     105.81  &       0.15  &       1.48  &       1.33  &      0 \\
                    l111w &     108.23  &      -0.43  &       3.35  &       2.78  &      0 \\
                    l111n &     110.50  &       2.18  &       4.19  &       2.21  &      0 \\
                    l111s &     111.07  &      -1.64  &       2.32  &       1.10  &      0 \\
                     l119 &     119.40  &       3.08  &       3.29  &       0.83  &    330 \\
                     l123 &     123.68  &       2.65  &       2.87  &       1.07  &     12 \\
                     l126 &     125.70  &       1.93  &       1.06  &       1.08  &      0 \\
                     l129 &     129.21  &       0.11  &       1.82  &       1.63  &      0 \\
                   camob1 &     141.20  &      -0.31  &       2.79  &       3.40  &      0 \\
                     l154 &     154.83  &       2.38  &       1.68  &       1.27  &      0 \\
                     l169 &     169.42  &      -0.32  &       4.08  &       2.05  &      0 \\
                    sh235 &     172.94  &       2.50  &       4.60  &       1.34  &      0 \\
                     l181 &     181.11  &       4.40  &       2.19  &       1.20  &      0 \\
                     l182 &     182.36  &       0.23  &       3.25  &       1.18  &     28 \\
                     l195 &     195.92  &      -0.66  &       3.04  &       1.18  &     56 \\
                     l201 &     201.57  &       0.30  &       1.32  &       1.37  &      0 \\
                  ngc2264 &     202.97  &       2.21  &       2.20  &       1.32  &      0 \\
              orionBnorth &     204.01  &     -11.86  &       2.17  &       1.33  &    335 \\
                   orionB &     206.73  &     -16.21  &       2.36  &       2.35  &     30 \\
              orionAspine &     212.45  &     -19.24  &       4.35  &       2.48  &      0 \\
                    monr2 &     213.54  &     -12.13  &       2.70  &       2.78  &      0 \\
                     l217 &     217.69  &      -0.24  &       1.91  &       1.04  &      0 \\
}
{\\
All numbers are in degrees.
}

The new target fields were selected from visual inspection of FCRAO OGS
\twelveco\ integrated maps, \citet{Dame2001} \twelveco\ maps, and IRAS 100 \um\
maps.  The fields were selected primarily to provide even spacing in RA in
order to maximize observing efficiency, and were therefore not blindly
selected.

Additionally, the Orion A and B and Mon R2 clouds were observed in observing
campaigns by collaborators.  These complexes are not directly part of the BGPS,
but are included in this data release reduced in the same manner as the
Galactic plane data.  They are much closer than typical BGPS sources and their
selection for mapping is very biased, but we include them in the archival data.
Parts of the Orion A nebula remain proprietary as of this release, but are
expected to be released upon publication of Kauffmann et al (in prep).  The
California nebula has also been observed and the data published in
\citet{Harvey2013a}.

Finally, some archival CSO data was recovered and added to the BGPS.  These
data include maps of M16 and M17.  M17 is an extraordinarily bright 1.1 mm
source that was poorly covered in the BGPS because it is below $b = -0.5$.

The Bolocat cataloging tool was run on these new fields and they have been
included in the \vtwo catalog.  Some of their properties are displayed in
Section \okinfinal{\ref{sec:sourceextraction}}.  A total of \nbolocatnew new
sources not covered in the \vone survey were extracted.\footnote{In the v2.0
catalog on IPAC, 35 sources in the $\ell=195$ and Orion B fields were inadvertently
excluded; these are now included in a v2.1 release.}

\section{Data Reduction and Data Products}
\label{sec:datareduction}

\subsection{A brief review of ground-based millimeter observational techniques}
Observations at wavelengths longer than 2 \um and shorter than 2 cm
from the ground are strongly affected by emission and absorption from our own
atmosphere.
Optical and radio observations from the ground see through a transparent
atmosphere, but millimeter observations are dominated by bright foreground
emission that dominates the astrophysical signal.  This foreground must be
removed in order to create maps of astrophysical emission.  

\citet{Chapin2013a} presented a summary of the techniques used to separate
astrophysical and atmospheric signals in (sub)millimeter bolometric
observations.  The Bolocam observations reported in this paper were conducted
with a fast-scanning strategy that places some of  the `fixed' astrophysical
emission at a different sampling frequency than the varying foreground
atmosphere.  This approach is one of the most efficient and flexible and has
been used predominantly over alternatives, such as a nodding secondary, in most
recent large-scale observing campaigns \citep{Aguirre2011,Schuller2012a}.

A variety of different atmospheric removal algorithms have been successfully
utilized, but in addition to removing the atmospheric foreground, these
approaches remove some of the astrophysical signal.  In order to recover signal
on angular scales up to the array size, the most commonly used approach for
bright Galactic signals is an iterative reconstruction process.  This process
assembles a model of the astrophysical emission, subtracts it from the observed
timestream, and repeats, each time reducing the amount of astrophysical signal
that is removed by the algorithm.  This general approach was first used on
Bolocam data by \citet{Enoch2006b} and refined in \citet{Aguirre2011}.  We
directly examine the effects of the data reduction methods below.

\subsection{Sky Subtraction}
We compared a few different methods for atmospheric subtraction and
astrophysical image reconstruction, but settled on an approach very similar to
that used in \vone.  This subsection recounts the minor changes from \vone and
includes discussion of alternative approaches.

The PCA method \citep{Enoch2006b,enoch07} with iterative flux density restoration was used
for \vtwo as for \vone.  In the PCA atmosphere removal method, the $n$
eigenvectors corresponding to the highest values along the diagonal of the
covariance matrix (the most correlated components) are nulled.  We nulled 13
PCA components in both \vone and \vtwo.  The selection of 13 components
produced the best compromise between uniform background noise and fully
restored peak signal.  Simulations show that the point source recovery is a
very weak function of number of PCA components nulled (nPCA), while extended
flux recovery is a strong function of nPCA.  However, residual atmospheric
signal was substantially reduced with higher nPCA.  In \vtwo, 20 iterations
were used instead of the 50 used in \vone; in both surveys, convergence was
clearly achieved by 20 iterations, and generally individual iterations are
indistinguishable by $\sim5$ iterations.

The iterative process adopts the non-negative flux density above some cutoff as a
model of the astrophysical sky and subtracts that flux density from the
timestream before repeating the atmospheric subtraction.  This approach allows
large angular scale structures to be recovered by removing them from the
timestreams before they can contribute to the correlated signal.  

The \vtwo pipeline was more succesful than in \vone at removing negative bowls
(see Section \ref{sec:catalogmatching} for visual examples).  Negative bowls
are introduced because the atmospheric subtraction process assumes that the
mean level of any timestream, and therefore any map, is zero.  The iterative
process allows this assumption to be violated, creating maps with net positive
signal.

The reduced impact of negative bowls is
attributed to a few small changes to the pipeline that each slightly mitigate
the bowls. 
\begin{enumerate}
    \item The astrophysical model is created by deconvolving the positive
emission rather using positive pixels directly.
The deconvolution process, which removes sub-beam-scale noise, was made more stable
in \vtwo by performing a local signal-to-noise cut using the noise maps
described in Section
\ref{sec:noisemaps}; in \vone there was no reliable noise map available during
the iterative map making process. 
    \item Better image co-alignment reduced inter-observation spatial offsets.  
        The astrophysical models therefore better reproduced the timestream data.
    \item Improvement in the bolometer gain calibration, which
is done on a per-observation basis in \vtwo, improved the convergence of the
iterative map maker.
\end{enumerate}
These changes are individually minor, but together resulted in significant
improvements to the map quality.

The quadratic planar fit sky subtraction method discussed in \citet{Sayers2010}
was implemented and tested for 1.1mm Galactic plane data in the v2.0 pipeline,
but was not used for the final data products.  In principle, this method should
do a substantially better job at removing smooth atmospheric signal from
timestreams than PCA cleaning because it is based on physically 
expected atmospheric variation.  The spatial recovery was better than the
aggressive 13-PCA approach, but as with a simpler median subtraction
approach (subtracting the median timestream from all bolometers), a great deal
of spurious signal from the atmosphere remained in the maps, and the noise
properties were highly non-uniform, rendering source extraction difficult.   It
was also more computationally expensive and did not remove correlated
electronic readout noise, which PCA subtraction did.  The \citet{Sayers2010} approach
is likely more effective at 143 GHz because the atmosphere is better-behaved at
lower frequencies.  We speculate that it is also more effective for deep
extragalactic fields in which more repeat observations of the same field are
able to distinguish atmospheric from real signal on the angular scales of the
array.

\subsection{Data products}
The BGPS data are available from the Infrared Processing and Analysis Center
(IPAC) at \url{irsa.ipac.caltech.edu/data/BOLOCAM_GPS/}.  The \vtwo data products
include the pipeline-processed maps and Bolocat label masks as in \vone.

In the \vtwo data release, there are two new map types released: noise maps and
median maps.  A variant of the noise maps was produced in \vone, while the
median maps are an entirely new data product.

\subsubsection{Noise Maps}
\label{sec:noisemaps}
Residual bolometer timestreams are automatically generated as part of the
iterative map-making process.  The residual is the result of subtracting the
astrophysical model (which is smooth, noiseless, and non-negative) from the
atmosphere-subtracted data timestream.  The resulting timestream should only
contain the remaining astrophysical noise.  However, maps of the residual
timestream contain sharp edge features because the astrophysical model is
sharp-edged (i.e., transitions from 0 to a non-zero value from one pixel to the
next).  These sharp transitions are mitigated in the presence of noise.

We therefore created noise maps by taking the local standard deviation of the
residual map.  Pixels in the original map that were not sampled (i.e.,
represented by \texttt{NaN} in the FITS data file) are ignored when computing
this local standard deviation and their values are set to be an arbitrarily
high number ($100$ Jy/beam) such that pixels near the map edge are assumed to
have extremely high noise (which is reasonable, since these pixels are affected
by a variety of artifacts rendering them unreliable measurements of the true
astrophysical flux).  The local noise is computed within a $FWHM=10$-pixel
gaussian, which enforces a high noise level within $\sim2\arcmin$ of the map
edge.  This method produces good noise maps (i.e., in agreement with the
standard deviation calculated from blank regions of the signal map) and was
used both within the iterative process and for cataloging.

We show the noise per pixel for each half square degree in the inner galaxy in
Figure \ref{fig:noiseperdeg}.  The noise level in each outer-galaxy field  is
summarized in Figure \ref{fig:noiseperfield}.  Because the outer galaxy
coverage is irregular, we show the noise per observed region rather than
dividing the regions into degree-scale sub-regions.

\Figure{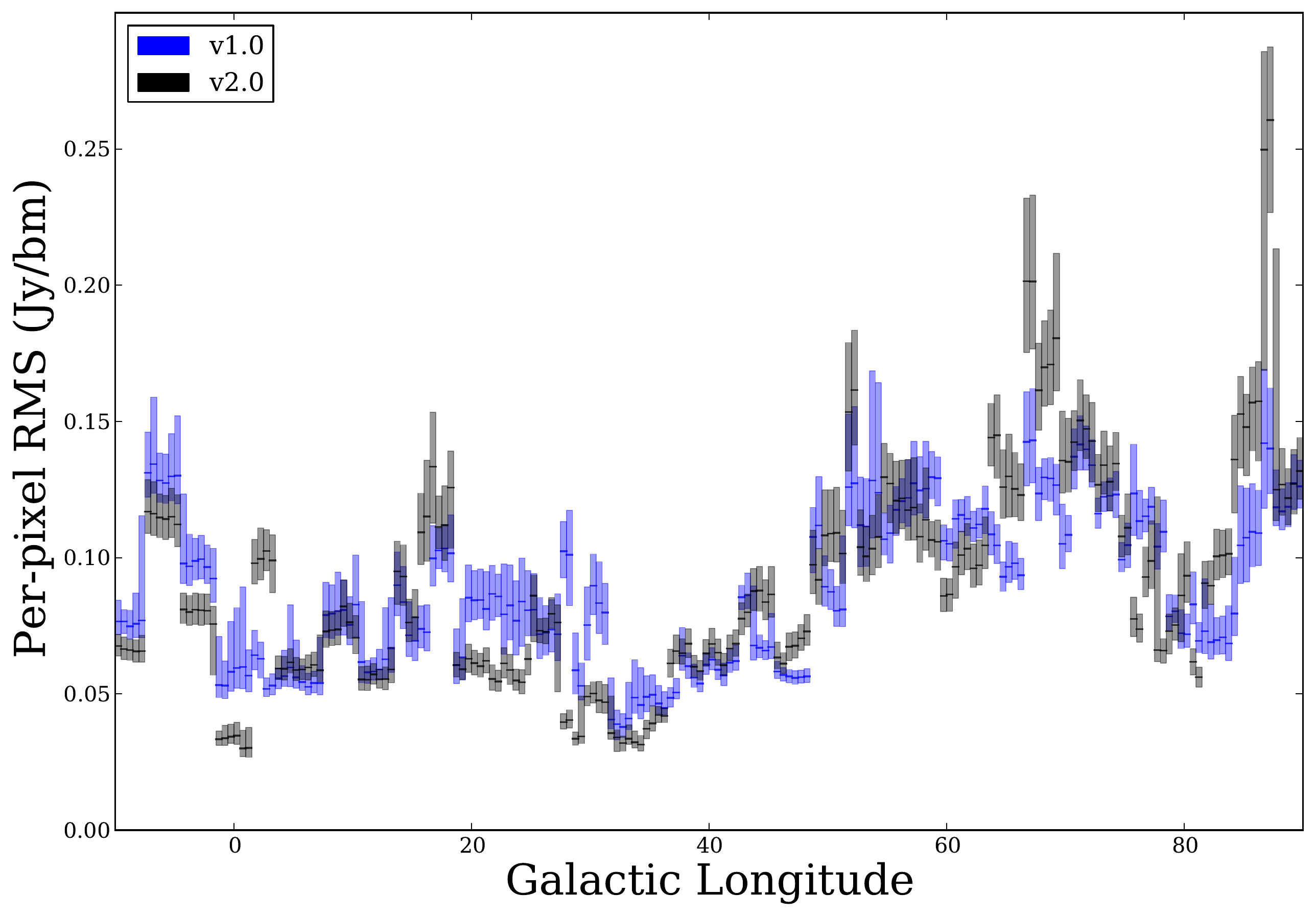}
{Map noise for 0.5 degree cuts in longitude in the range $|b|<0.5 \degrees$.  The solid
horizontal lines show the median noise in the map, while the shaded regions
highlight the 1-$\sigma$ (68\%) interval (quantiles 16-84) of the noise.  The
noise is the local weighted standard deviation (rms) over a $FWHM=10$-pixel
region (see Section \ref{sec:noisemaps}).
}
{fig:noiseperdeg}{0.5}{0}

\Figure{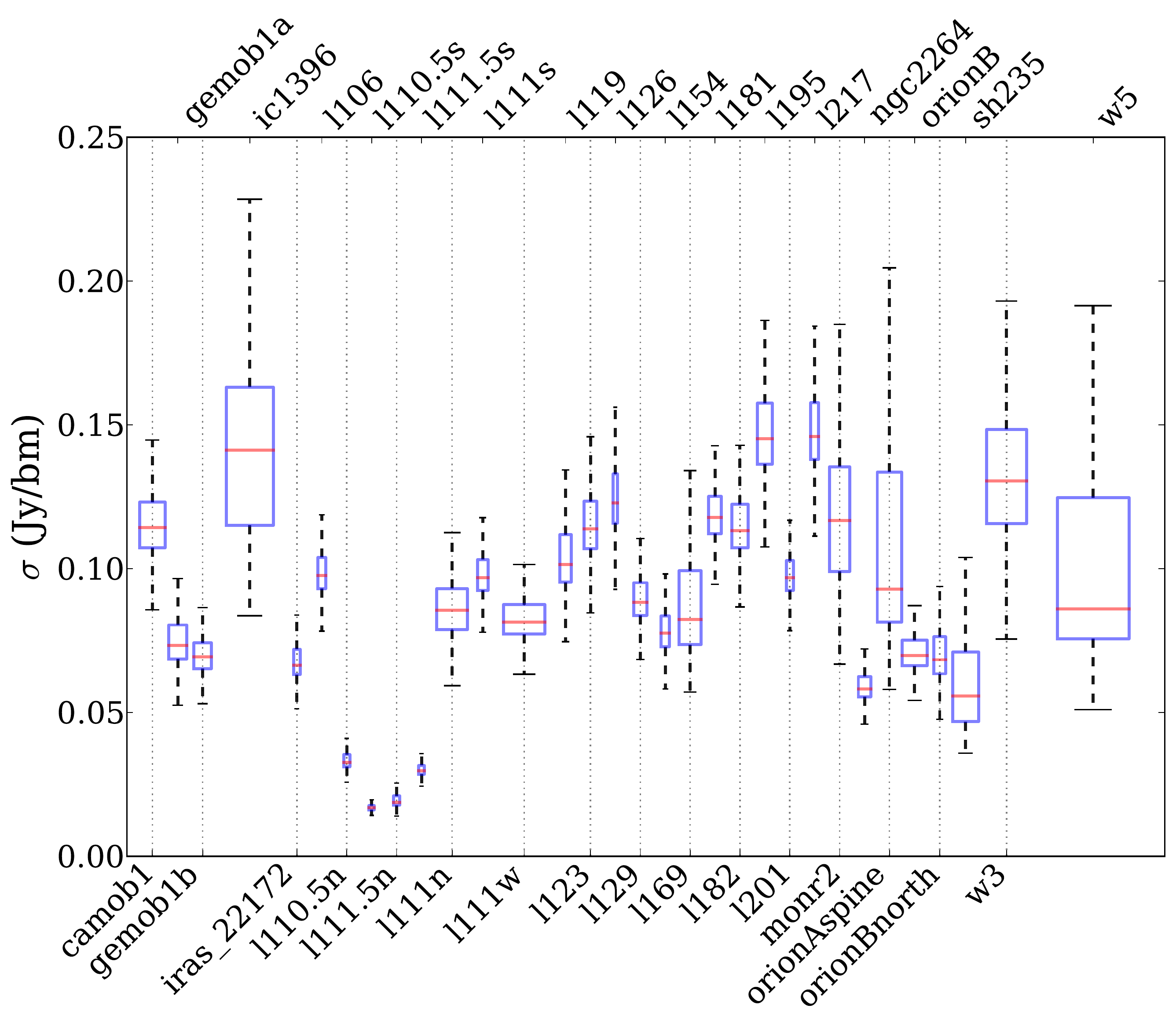}
{The noise in each outer galaxy field shown with box plots.  The red lines
indicate the median, the boxes show the 25\%-75\% range, and the black dashed
lines (`whiskers') show the 16\%-84\% (1-$\sigma$) range.  Unlike Figure
\ref{fig:noiseperdeg}, the field size for each region varies, which is why
there is a much broader spread in the widths of the individual noise
distributions.  The size of the region is proportional to the box width.
}{fig:noiseperfield}{0.5}{0}

\subsection{Median Maps}

Some artifacts (cosmic ray hits, instrumental artifacts) inevitably remained at
the end of the process.  In order to mitigate these effects, ``median maps''
were created.  The value of each spatial pixel was set to the median value of
the timestream points intersecting that pixel; pixels with fewer than 3 data points
were set to \texttt{NaN}.  The noise in the median
maps was in some cases lower than that in the weighted mean maps, particularly for
fields with fewer total observations.  They uniformly have mitigated
instrument-related artifacts such as streaking.  These maps are released in
addition to the weighted-mean maps, which often have higher signal-to-noise.

\subsection{Pointing}
\label{sec:pointing}
In order to get the best possible pointing accuracy in each field, all
observations of a given area were median-combined using the {\sc montage}
package, which performs image reprojections, to create a pointing master map \citep{Berriman2004a}.
Each individual observation was then aligned to the master using a
cross-correlation technique \citep{Welsch2004a}:
\begin{enumerate}
    \item The master and
target image were projected to the same pixel space 
    \item A cross-correlation image was generated, and the peak pixel in the
        cross-correlation map was identified
    \item Sub-pixel alignment was measured by performing a 2nd-order Taylor
        expansion around the peak pixel
\end{enumerate}
        
This method is similar to the version 1.0 method, but the new peak-finding
method proved more robust than the previous Gaussian fitting approach.  The
\vone Gaussian fitting approach is often used
in astronomy (e.g.,
\url{http://www.astro.ucla.edu/~mperrin/IDL/sources/subreg.pro}), but it is biased
when images are dominated by extended structure.  This bias occurs because the least-squares
fitting approach will identify the broader peak that represents
auto-correlation of astrophysical structure rather than the cross-correlation between
the two images.  In \vone, we attempted to mitigate this issue by subtracting
off a `background' component before fitting the Gaussian peak, but this method
was not robust.

The improved approach to pointing resulted in typical RMS offsets between
the individual frames and the master $\sigma_{\mathbf{x}} \sim 2$ \arcsec.   The
improvement in the point spread function is readily observed
(see Section \ref{sec:catalogmatching}).

\subsection{Pointing Comparison}
We carefully re-examined the pointing throughout the BGPS using a
degree-by-degree cross-correlation analysis between the \vone, \vtwo, and
Herschel Hi-Gal 350 \um data.  The Herschel data were unsharp-masked (high-pass filtered) by subtracting a
version of the data smoothed with a $\sigma=120\arcsec$ Gaussian.  The result
was then convolved with a $\sigma=8.9\arcsec$ Gaussian to match the Herschel
to the Bolocam beam sizes.

Errors on the offsets were measured utilizing the
Fourier scaling theorem to achieve sub-pixel resolution \citep[inspired by
][]{Guizar2008}.  The errors on the best-fit shift were determined 
using errors estimated from the BGPS data and treating the filtered
Hi-Gal data as an ideal (noiseless) model.  The tools for this process, along with a test
suite demonstrating their applicability to extended structures in images, are
publicly available at \url{http://image-registration.readthedocs.org/}.

The cross-correlation technique calculated the $\chi^2$ statistic as a function
of the offset.  For a reference image $Y$ and observed image $X$ with error per
pixel $\sigma_{xy}$, $$\chi^2 = \sum \frac{(X-Y(\Delta x,\Delta y))^2}{\sigma_{xy}^2}$$ where $\Delta x$ and $\Delta y$
are the pixel shifts.  Because Y is not actually an ideal model but instead is
a noisy image, we increase $\sigma_{xy}$ by the rms of the difference between the
aligned images, using a corrected $\sigma_c^2 = \sigma_{BGPS}^2 + RMS(X-Y(\Delta x_b,\Delta y_b))$, where
$\Delta x_b,\Delta y_b$ are the best-fit shifts.

For the majority of the examined 1-square-degree fields, the signal dominated the noise
and we were able to measure the offsets to sub-pixel accuracy.
A plot of the longitude / latitude offsets between \vtwo\ and \vone\ and
Herschel Hi-Gal is shown in Figure \ref{fig:offsets}.  

Table \ref{tab:ccoffsets} lists the measured offsets in arcseconds between images
for all 1 degree fields from $\ell=351\arcdeg$ to $\ell=65\arcdeg$.  The offsets represent
the Galactic longitude and latitude shifts in arcseconds from the reference (left) to the
`measured' field (right).  

Table \ref{tab:ccoffsetsmeans} shows the means of the columns in Table
\ref{tab:ccoffsets}, weighted by the error in the measurements and by the
number of sources.  Weighting by the number of sources is used for comparison
with other works that attempt to measure the pointing offset on the basis of
catalog source position offsets.  None of the measured offsets are significant;
in all cases the scatter exceeds the measured offset.

\TallFigureTwo{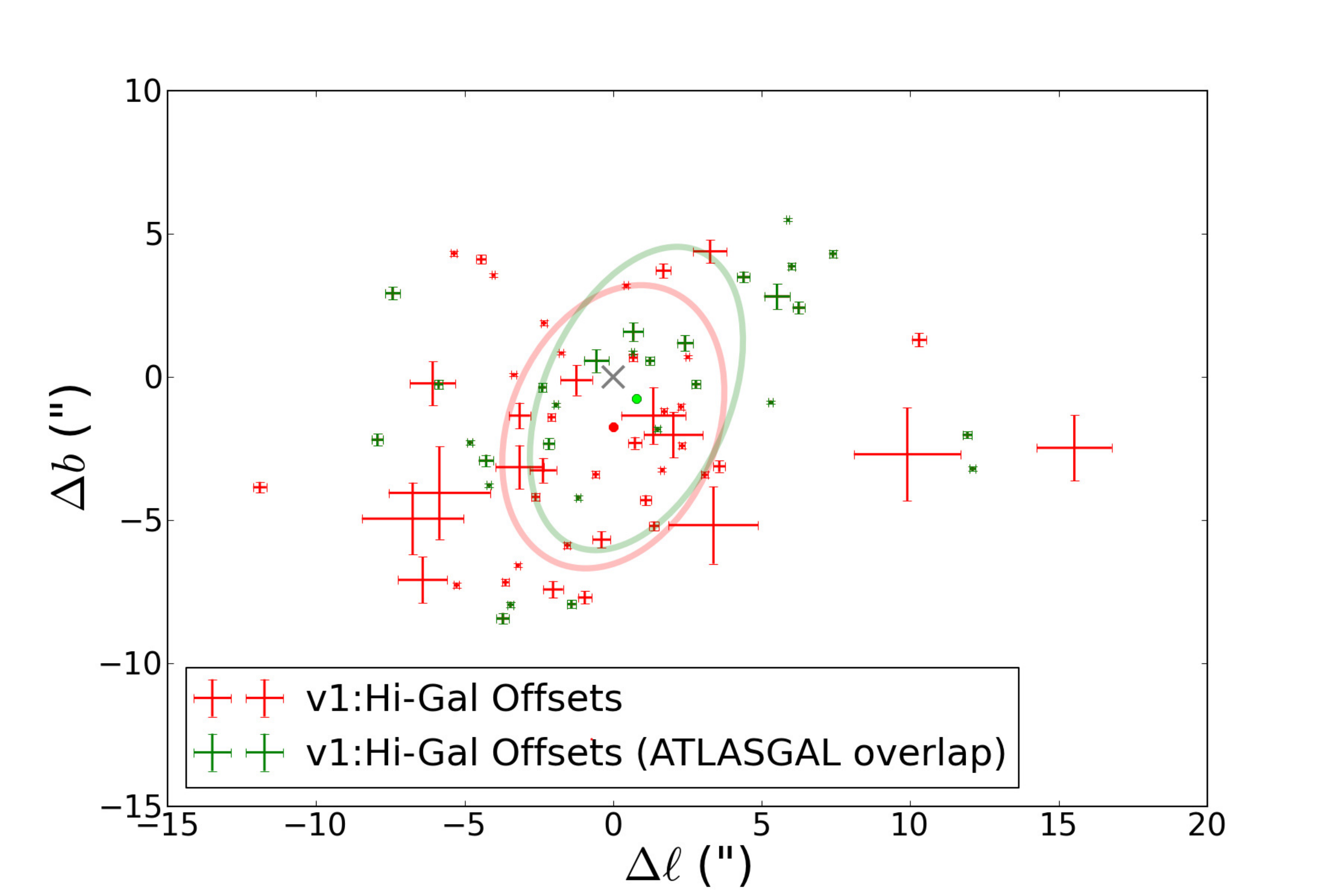}
{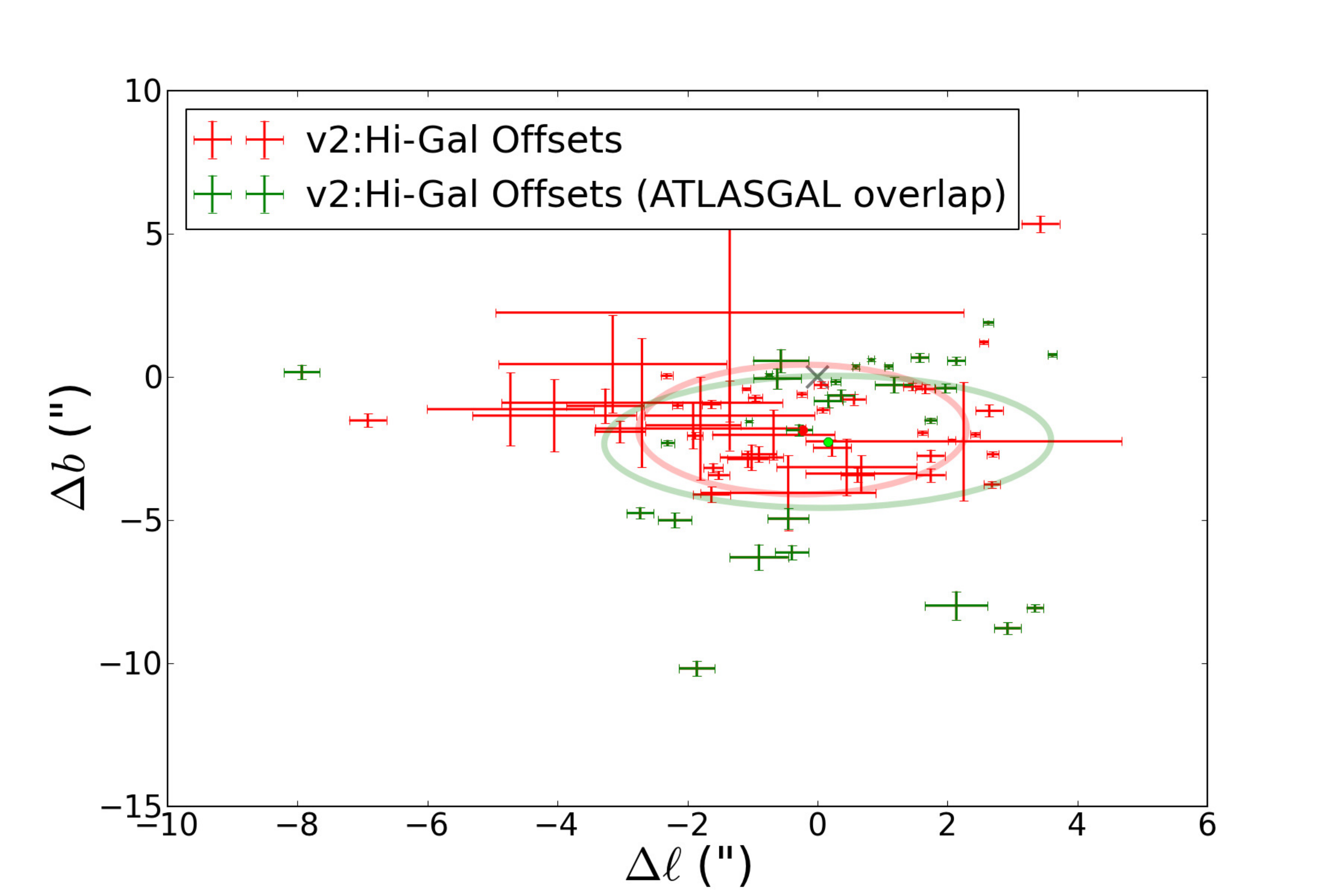}
{Plots of the latitude and longitude offsets of individual 1-degree fields in
\vone (a) and \vtwo (b) as compared with Herschel Hi-Gal.  Offsets were measured using
a cross-correlation technique described in the text.  The error bars correspond
to $\Delta \chi^2 < 2.3$, or $1-\sigma$ for Gaussian distributed noise and 2
degrees of freedom.  The circles and ellipses represent the mean and standard
deviation (unweighted) offsets in the whole survey (red) and the $(351\arcdeg < \ell)
\cup (\ell < 20\arcdeg)$ ATLASGAL-overlap regions (green).   In both cases, the mean
offset is consistent with zero (shown as a black \texttt{x}), but many individual fields show significant
offsets.  Note that the scales are different; there are far fewer outliers
in the \vtwo-Herschel comparison (b) and the average offset is much closer to zero.
The errors are larger in the non-ATLASGAL overlap region because there is less
signal in the $35\arcdeg<\ell<65\arcdeg$ range.
}
{fig:offsets}{5in}{5in}

\Table{lccccccc}{Cross-Correlation Offsets}
{\colhead{     Field Name}&\colhead{$\Delta\ell$(HG-v2)}&\colhead{$\Delta b$(HG-v2)}&\colhead{$\Delta\ell$(v1-v2)}&\colhead{$\Delta b$(v1-v2)}&\colhead{$\Delta\ell$(HG-v1)}&\colhead{$\Delta b$(HG-v1)}&\colhead{  N(v1 sources)}}
{tab:ccoffsets}{
           l351 &  $0.37 (-0.21)$ &  $-0.65 (0.21)$ &  $3.12 (-0.16)$ &   $1.83 (0.14)$ & $-2.17 (-0.19)$ &  $-2.33 (0.19)$ &              56 \\
           l352 &  $1.10 (-0.07)$ &   $0.37 (0.07)$ &  $3.14 (-0.06)$ &   $1.45 (0.06)$ & $-1.93 (-0.07)$ &  $-0.97 (0.07)$ &              87 \\
           l353 &  $3.35 (-0.13)$ &  $-8.07 (0.13)$ & $-3.80 (-0.08)$ & $-12.41 (0.08)$ &  $7.39 (-0.12)$ &   $4.30 (0.12)$ &              65 \\
           l354 &  $2.14 (-0.48)$ &  $-7.98 (0.50)$ & $-3.71 (-0.11)$ & $-12.04 (0.09)$ &  $5.51 (-0.43)$ &   $2.81 (0.44)$ &              52 \\
           l355 &  $2.92 (-0.20)$ &  $-8.78 (0.20)$ & $-3.57 (-0.11)$ & $-12.12 (0.09)$ &  $6.24 (-0.20)$ &   $2.42 (0.20)$ &              54 \\
           l356 & $-0.62 (-0.37)$ &  $-0.06 (0.35)$ & $-1.29 (-0.22)$ &  $-0.96 (0.17)$ &  $0.68 (-0.34)$ &   $1.58 (0.32)$ &              42 \\
           l357 & $-0.56 (-0.43)$ &   $0.56 (0.41)$ & $-0.06 (-0.24)$ &  $-0.39 (0.17)$ & $-0.56 (-0.42)$ &   $0.56 (0.39)$ &              23 \\
           l358 &  $2.14 (-0.14)$ &   $0.56 (0.14)$ & $-0.62 (-0.14)$ &   $0.17 (0.12)$ &  $2.78 (-0.14)$ &  $-0.25 (0.14)$ &              35 \\
           l359 &  $2.63 (-0.08)$ &   $1.90 (0.06)$ & $-10.05 (-0.08)$ &   $5.86 (0.07)$ & $12.10 (-0.10)$ &  $-3.21 (0.07)$ &             248 \\
           l000 &  $3.61 (-0.07)$ &   $0.77 (0.06)$ & $-1.10 (-0.04)$ &   $1.55 (0.03)$ &  $5.30 (-0.07)$ &  $-0.89 (0.06)$ &             318 \\
           l001 &  $0.59 (-0.05)$ &   $0.37 (0.07)$ & $-3.36 (-0.05)$ &   $1.31 (0.06)$ & $11.92 (-0.16)$ &  $-2.02 (0.11)$ &             368 \\
           l002 & $-0.39 (-0.26)$ &  $-6.13 (0.25)$ &  $1.46 (-0.24)$ &   $2.14 (0.19)$ & $-1.41 (-0.15)$ &  $-7.93 (0.14)$ &             170 \\
           l003 & $-2.73 (-0.21)$ &  $-4.75 (0.20)$ &  $0.62 (-0.18)$ &   $3.54 (0.17)$ & $-3.46 (-0.10)$ &  $-7.96 (0.10)$ &             243 \\
           l004 & $-1.86 (-0.28)$ & $-10.18 (0.26)$ & $-0.34 (-0.23)$ &   $0.11 (0.19)$ & $-3.71 (-0.21)$ &  $-8.44 (0.19)$ &              70 \\
           l005 &  $1.18 (-0.29)$ &  $-0.28 (0.28)$ & $-3.99 (-0.19)$ &  $-3.54 (0.19)$ &  $4.39 (-0.21)$ &   $3.49 (0.19)$ &              78 \\
           l006 &  $0.83 (-0.05)$ &   $0.60 (0.06)$ & $-1.17 (-0.05)$ &  $-1.36 (0.06)$ &  $5.88 (-0.05)$ &   $5.48 (0.05)$ &             109 \\
           l007 &  $1.97 (-0.17)$ &  $-0.39 (0.16)$ & $-4.50 (-0.09)$ &  $-4.50 (0.00)$ &  $6.00 (-0.12)$ &   $3.87 (0.11)$ &              93 \\
           l008 &  $1.58 (-0.14)$ &   $0.68 (0.16)$ &  $4.58 (-0.12)$ &   $1.43 (0.15)$ & $-2.39 (-0.13)$ &  $-0.37 (0.15)$ &              59 \\
           l009 &  $1.74 (-0.10)$ &  $-1.52 (0.09)$ &  $6.15 (-0.09)$ &   $1.00 (0.09)$ & $-4.82 (-0.08)$ &  $-2.29 (0.08)$ &              55 \\
           l010 &  $0.28 (-0.08)$ &  $-0.17 (0.08)$ &  $3.70 (-0.08)$ &   $1.98 (0.09)$ & $-4.18 (-0.08)$ &  $-3.78 (0.07)$ &              77 \\
           l011 & $-1.05 (-0.04)$ &  $-1.56 (0.04)$ & $-0.94 (-0.03)$ &  $-0.04 (0.02)$ & $-1.18 (-0.07)$ &  $-4.22 (0.08)$ &             122 \\
           l012 & $-2.31 (-0.10)$ &  $-2.31 (0.10)$ & $-3.80 (-0.08)$ &  $-0.70 (0.09)$ &  $1.49 (-0.09)$ &  $-1.83 (0.08)$ &             102 \\
           l013 & $-0.75 (-0.05)$ &   $0.07 (0.05)$ & $-1.49 (-0.05)$ &  $-0.59 (0.05)$ &  $0.66 (-0.05)$ &   $0.86 (0.06)$ &             198 \\
           l014 & $-0.28 (-0.20)$ &  $-1.86 (0.20)$ &  $6.05 (-0.15)$ &  $-1.72 (0.16)$ & $-5.88 (-0.15)$ &  $-0.25 (0.15)$ &             137 \\
           l015 & $-2.19 (-0.26)$ &  $-5.01 (0.26)$ &  $5.79 (-0.17)$ &  $-2.64 (0.19)$ & $-7.93 (-0.20)$ &  $-2.19 (0.20)$ &             164 \\
           l016 & $-0.90 (-0.45)$ &  $-6.30 (0.45)$ &  $5.79 (-0.26)$ &  $-2.31 (0.28)$ & $-4.28 (-0.25)$ &  $-2.92 (0.20)$ &              63 \\
           l017 & $-0.45 (-0.32)$ &  $-4.95 (0.36)$ & $-1.29 (-0.22)$ &  $-3.88 (0.26)$ &  $2.42 (-0.26)$ &   $1.18 (0.27)$ &              62 \\
           l018 &  $0.17 (-0.23)$ &  $-0.84 (0.21)$ & $-1.07 (-0.17)$ &  $-1.97 (0.17)$ &  $1.24 (-0.15)$ &   $0.56 (0.14)$ &              55 \\
           l019 & $-7.93 (-0.28)$ &   $0.17 (0.25)$ &  $0.89 (-0.11)$ &  $-2.69 (0.10)$ & $-7.42 (-0.25)$ &   $2.92 (0.23)$ &             179 \\
           l020 &  $2.43 (-0.07)$ &  $-2.01 (0.08)$ &  $0.07 (-0.07)$ &  $-2.60 (0.06)$ &  $2.50 (-0.06)$ &   $0.70 (0.06)$ &             110 \\
           l021 &  $2.64 (-0.21)$ &  $-1.18 (0.21)$ &  $1.86 (-0.17)$ &  $-1.74 (0.16)$ &  $0.68 (-0.14)$ &   $0.68 (0.14)$ &             103 \\
           l022 &  $1.74 (-0.21)$ &  $-2.76 (0.21)$ & $-0.65 (-0.16)$ &   $0.48 (0.14)$ &  $3.57 (-0.20)$ &  $-3.12 (0.20)$ &              87 \\
           l023 &  $2.69 (-0.12)$ &  $-3.75 (0.11)$ & $-0.20 (-0.09)$ &  $-0.31 (0.08)$ &  $3.08 (-0.10)$ &  $-3.42 (0.10)$ &             213 \\
           l024 &  $2.70 (-0.09)$ &  $-2.70 (0.09)$ & $-0.44 (-0.09)$ &  $-0.27 (0.08)$ &  $2.32 (-0.10)$ &  $-2.40 (0.10)$ &             250 \\
           l025 &  $1.62 (-0.08)$ &  $-1.95 (0.08)$ &  $0.08 (-0.07)$ &   $1.15 (0.07)$ &  $1.65 (-0.06)$ &  $-3.25 (0.07)$ &             183 \\
           l026 &  $1.66 (-0.15)$ &  $-0.42 (0.16)$ &  $0.06 (-0.12)$ &   $0.62 (0.13)$ &  $1.72 (-0.10)$ &  $-1.21 (0.11)$ &             151 \\
           l027 &  $1.46 (-0.15)$ &  $-0.34 (0.14)$ & $-0.48 (-0.12)$ &   $0.87 (0.12)$ &  $2.28 (-0.10)$ &  $-1.04 (0.10)$ &             119 \\
           l028 &  $2.56 (-0.07)$ &   $1.21 (0.08)$ &  $6.22 (-0.07)$ &   $8.35 (0.07)$ & $-3.63 (-0.11)$ &  $-7.17 (0.12)$ &             188 \\
           l029 & $-0.96 (-0.11)$ &  $-0.73 (0.11)$ &  $4.58 (-0.09)$ &   $6.38 (0.09)$ & $-5.27 (-0.09)$ &  $-7.27 (0.09)$ &             177 \\
           l030 &  $0.06 (-0.11)$ &  $-0.28 (0.12)$ &  $2.05 (-0.09)$ &   $3.23 (0.11)$ & $-3.21 (-0.07)$ &  $-6.58 (0.06)$ &             276 \\
           l031 & $-1.10 (-0.06)$ &  $-0.42 (0.05)$ & $-1.69 (-0.03)$ &  $-1.46 (0.02)$ &  $0.44 (-0.07)$ &   $3.19 (0.07)$ &             354 \\
           l032 & $-0.24 (-0.08)$ &  $-0.60 (0.09)$ &  $1.18 (-0.08)$ &  $-2.42 (0.08)$ & $-2.33 (-0.09)$ &   $1.88 (0.08)$ &             189 \\
           l033 &  $2.07 (-0.05)$ &  $-2.21 (0.05)$ &  $6.10 (-0.06)$ &  $-5.71 (0.06)$ & $-4.04 (-0.06)$ &   $3.56 (0.06)$ &             210 \\
           l034 & $-2.32 (-0.09)$ &   $0.04 (0.08)$ & $-0.46 (-0.05)$ &  $-0.10 (0.05)$ & $-1.55 (-0.11)$ &  $-5.88 (0.10)$ &             203 \\
           l035 & $-1.88 (-0.12)$ &  $-2.05 (0.12)$ &  $0.31 (-0.11)$ &  $-0.59 (0.10)$ & $-2.08 (-0.13)$ &  $-1.41 (0.12)$ &             247 \\
           l036 & $-1.63 (-0.14)$ &  $-0.96 (0.15)$ &  $0.82 (-0.12)$ &   $3.01 (0.11)$ & $-2.62 (-0.14)$ &  $-4.19 (0.14)$ &             126 \\
           l037 & $-1.07 (-0.32)$ &  $-2.87 (0.29)$ & $-2.31 (-0.25)$ &   $0.51 (0.24)$ &  $0.73 (-0.23)$ &  $-2.31 (0.21)$ &              83 \\
           l038 & $-1.60 (-0.15)$ &  $-3.18 (0.17)$ & $-0.79 (-0.12)$ &   $0.34 (0.14)$ & $-0.59 (-0.11)$ &  $-3.40 (0.12)$ &              69 \\
           l039 &  $0.62 (-0.26)$ &  $-3.43 (0.24)$ & $-0.23 (-0.16)$ &   $1.12 (0.14)$ &  $1.10 (-0.19)$ &  $-4.30 (0.17)$ &              69 \\
           l040 &  $1.74 (-0.23)$ &  $-3.43 (0.24)$ &  $1.86 (-0.18)$ &   $2.98 (0.17)$ &  $1.38 (-0.17)$ &  $-5.20 (0.16)$ &              40 \\
           l041 &  $0.23 (-0.29)$ &  $-2.48 (0.29)$ &  $1.07 (-0.24)$ &   $5.23 (0.20)$ & $-0.96 (-0.23)$ &  $-7.70 (0.23)$ &              44 \\
           l042 & $-1.01 (-0.48)$ &  $-2.81 (0.45)$ &  $0.11 (-0.26)$ &   $5.06 (0.23)$ & $-2.02 (-0.34)$ &  $-7.42 (0.29)$ &              36 \\
           l043 &  $0.08 (-0.10)$ &  $-1.15 (0.09)$ &  $2.07 (-0.08)$ &  $-2.52 (0.09)$ & $-1.74 (-0.07)$ &   $0.84 (0.05)$ &              17 \\
           l044 & $-1.63 (-0.29)$ &  $-4.11 (0.27)$ &  $4.78 (-0.26)$ &  $-8.16 (0.20)$ & $-4.44 (-0.17)$ &   $4.11 (0.15)$ &              27 \\
           l045 & $-1.52 (-0.16)$ &  $-3.43 (0.15)$ &  $4.42 (-0.14)$ &  $-7.73 (0.12)$ & $-5.36 (-0.10)$ &   $4.32 (0.09)$ &              30 \\
           l046 & $-0.90 (-0.27)$ &  $-2.70 (0.27)$ & $-1.24 (-0.26)$ &  $-5.29 (0.21)$ &  $1.69 (-0.25)$ &   $3.71 (0.25)$ &              53 \\
           l047 &  $0.68 (-0.85)$ &  $-3.38 (0.63)$ &  $0.34 (-0.38)$ &  $-3.71 (0.27)$ &  $3.26 (-0.56)$ &   $4.39 (0.41)$ &              11 \\
           l048 & $-0.45 (-1.35)$ &  $-4.05 (1.31)$ &  $8.16 (-0.27)$ &  $-3.32 (0.30)$ & $-6.08 (-0.76)$ &  $-0.23 (0.76)$ &               6 \\
           l049 & $-2.15 (-0.08)$ &  $-1.00 (0.08)$ & $-0.35 (-0.05)$ &  $-0.75 (0.06)$ & $-3.35 (-0.09)$ &   $0.08 (0.05)$ &             113 \\
           l050 &  $0.56 (-0.18)$ &  $-0.79 (0.19)$ &  $1.77 (-0.14)$ &  $-0.14 (0.12)$ & $-0.73 (-0.06)$ & $-12.66 (0.30)$ &              31 \\
           l051 & $-1.91 (-0.73)$ &  $-1.69 (0.82)$ &  $4.95 (-0.23)$ &   $0.45 (0.32)$ & $-3.15 (-0.36)$ &  $-1.35 (0.45)$ &               9 \\
           l052 & $-2.70 (-2.16)$ &  $-0.90 (2.25)$ & $-3.71 (-0.54)$ &   $3.26 (0.21)$ &  $3.38 (-1.51)$ &  $-5.18 (1.35)$ &               0 \\
           l053 & $-3.04 (-0.39)$ &  $-1.91 (0.38)$ & $-1.01 (-0.28)$ &   $2.59 (0.19)$ & $-0.39 (-0.30)$ &  $-5.68 (0.28)$ &              26 \\
           l054 & $-3.26 (-0.60)$ &  $-1.01 (0.61)$ &  $1.35 (-0.27)$ &   $1.35 (0.18)$ & $-2.36 (-0.45)$ &  $-3.26 (0.43)$ &              26 \\
           l055 & $-4.05 (-1.26)$ &  $-1.35 (1.26)$ & $-0.23 (-0.34)$ &   $1.58 (0.23)$ & $-3.15 (-0.81)$ &  $-3.15 (0.76)$ &               4 \\
           l056 & $-4.72 (-1.28)$ &  $-1.12 (1.28)$ &  $4.39 (-0.44)$ &  $-0.79 (0.24)$ & $-1.24 (-0.53)$ &  $-0.11 (0.53)$ &              10 \\
           l057 & $-3.15 (-1.76)$ &   $0.45 (1.71)$ &  $2.81 (-0.50)$ &   $0.11 (0.21)$ &  $1.35 (-1.08)$ &  $-1.35 (0.99)$ &               1 \\
           l058 & $-1.35 (-1.30)$ &  $-1.35 (1.22)$ & $-6.86 (-0.48)$ &   $0.79 (0.25)$ &  $2.02 (-0.99)$ &  $-2.02 (0.79)$ &               4 \\
           l059 &  $0.45 (-1.08)$ &  $-3.15 (0.99)$ & $-4.28 (-0.23)$ &   $1.12 (0.68)$ & $-6.75 (-1.71)$ &  $-4.95 (1.26)$ &               2 \\
           l060 &  $3.43 (-0.29)$ &   $5.34 (0.29)$ & $-7.09 (-0.14)$ &   $4.61 (0.17)$ & $10.30 (-0.24)$ &   $1.29 (0.23)$ &              17 \\
           l061 & $-6.92 (-0.29)$ &  $-1.52 (0.23)$ &  $5.12 (-0.16)$ &   $3.54 (0.16)$ & $-11.89 (-0.23)$ &  $-3.85 (0.19)$ &               4 \\
           l062 & $-1.80 (-1.62)$ &  $-1.80 (1.80)$ &  $4.72 (-0.18)$ &   $5.18 (0.25)$ & $-5.85 (-1.71)$ &  $-4.05 (1.62)$ &               1 \\
           l063 & $-0.68 (-0.95)$ &  $-2.02 (0.88)$ &  $2.64 (-0.18)$ &   $4.67 (0.27)$ & $-6.41 (-0.83)$ &  $-7.09 (0.81)$ &               5 \\
           l064 &  $2.25 (-2.43)$ &  $-2.25 (2.07)$ & $-24.98 (-0.43)$ &  $-2.92 (0.83)$ & $15.52 (-1.26)$ &  $-2.48 (1.15)$ &               1 \\
           l065 & $-1.35 (-3.60)$ &   $2.25 (3.82)$ & $-24.08 (-0.63)$ &  $-4.28 (0.99)$ &  $9.90 (-1.80)$ &  $-2.70 (1.62)$ &               1 \\
}{\\
The offsets reported are in units of arcseconds, and the values in parentheses represent the 1-$\sigma$ error bars.}
\Table{lcccccc}{Cross-Correlation Offset Means and Standard Deviations}
{               &\colhead{$\Delta\ell$(HG-v2)}&\colhead{$\Delta b$(HG-v2)}&\colhead{$\Delta\ell$(v1-v2)}&\colhead{$\Delta b$(v1-v2)}&\colhead{$\Delta\ell$(HG-v1)}&\colhead{$\Delta b$(HG-v1)}}
{tab:ccoffsetsmeans}{
Mean & $           0.23$ & $           -1.8$ &$           0.16$ & $          -0.37$ &$        -0.0047$ & $           -1.7$ \\
Standard Deviation & $            2.2$ & $            2.5$ &$            5.3$ & $            3.9$ &$            4.9$ & $            3.7$ \\
Weighted Mean & $          -0.47$ & $          -0.89$ &$           0.26$ & $           -1.1$ &$         -0.089$ & $          -0.87$ \\
Weighted Standard Deviation & $            1.7$ & $            1.8$ &$            3.1$ & $            3.5$ &$              4$ & $            3.6$ \\
N(src) Weighted Mean & $          -0.24$ & $           -1.1$ &$           0.26$ & $           0.58$ &$           -1.3$ & $           -1.9$ \\
N(src) Weighted Standard Deviation & $            2.3$ & $            2.1$ &$            3.5$ & $            3.2$ &$            5.7$ & $            3.6$ \\
N(src) Weighted Mean $\ell<21$ & $          -0.24$ & $             -1$ &$            1.5$ & $           0.84$ &$           -3.7$ & $           -1.9$ \\
N(src) Weighted Standard Deviation $\ell<21$ & $            2.9$ & $            2.8$ &$            4.1$ & $            3.2$ &$            7.3$ & $            3.4$ \\
}{\\ 
The offsets reported are in units of arcseconds, and the values in parentheses represent the 1-$\sigma$ error bars.}

\subsection{Addressing the ATLASGAL offset}
\citet{Contreras2012} performed a comparison of the Bolocam and ATLASGAL
catalogs, identifying a systematic offset between the catalogs of
$\Delta\ell=-4.7\arcsec$, $\Delta b = 1.2 \arcsec$.  Because the offset is measured
between catalog points, the meaning of this measured offset is not immediately
clear.  In the BGPS maps in the ATLASGAL-BGPS overlap region, there were 12
individual sub-regions ($3\arcdeg\times1\arcdeg$, with $1\arcdeg\times1\arcdeg$
regions in the CMZ) that could have independendent pointing.  Because we did
not have direct access to the ATLASGAL maps or catalog at the time of writing,
we compared the Bolocam \vone and \vtwo catalogs to each other determine
whether the pointing changes in \vtwo might account for the observed ATLASGAL
offset, assuming the \vtwo pointing is more accurate than the \vone
pointing.  

We performed an inter-catalog match between \vone and \vtwo, considering sources
between the two catalogs to be a match if the distance between the centroid
positions of the two sources is $<40\arcsec$ (this distance is more conservative than that
used in Section
\okinfinal{\ref{sec:catalogmatching}}).  We then compared the pointing offset
as measured by the mean offset between the catalogs to the offset measured via
cross-correlation analysis of the maps on a per-square-degree basis.  The
catalog and image offsets agree well, with no clear systematic offsets between
the two estimators.  The scatter in the catalog-based measurements is
much greater, which is expected since the source positions are subject to spatial
scale recovery differences between the versions and because the sources include less
signal than the complete maps.

There is no clear net offset between either version of the BGPS and the
Herschel Hi-Gal survey, or between the two versions of the BGPS.  However, the
scatter in the pointing offsets between \vone and Herschel is substantially
greater than the \vtwo-Herschel offsets.  The offset measured in
\citet{Contreras2012} is likely a result of particularly large offsets in a few
fields with more identified sources.  As shown in Table
\ref{tab:ccoffsetsmeans}, the mean offset, \emph{weighted by number of sources}, is
greater for the ATLASGAL overlap region than overall.  We reproduce
a number similar to the ATLASGAL-measured longitude offset of
$\Delta\ell=-4.7\arcsec$ (our source-count-weighted
$\Delta\ell=-3.7\arcsec$), despite a much larger standard deviation and despite
no significant offset being measured directly in the images.
These measurements imply that the pointing offset measured by \citet{Contreras2012} 
was localized to a few fields \emph{and} that the offset is corrected in the \vtwo data.

\section{The Angular Transfer Function of the BGPS} 
\label{sec:stf}

\subsection{Simulations with synthetic sky and atmosphere}
\label{sec:atmotests}
In order to determine the angular response of the Bolocam array and BGPS
pipeline in realistic observing conditions, we performed simulations of a
plausible synthetic astrophysical sky with synthetic atmospheric signal added
to the bolometer timestream.

To generate the simulated atmosphere, we fit a piecewise power law to a power
spectrum of a raw observed timestream (Figure \ref{fig:powerspecfit}).  The
power spectrum varies in amplitude depending on weather conditions and
observation length, but the shape is generally well-represented by $1/f$
``pink'' noise ($P_\nu\propto \nu^{-1.5}$) for $\nu < 2 $ Hz and flat ``white''
noise ($P_\nu \sim const$) for $\nu \geq 2$ Hz, where $\nu$ is the frequency.
We show a fitted timestream power spectrum in Figure \ref{fig:powerspecfit}.
The deviations from $1/f$ and white noise have little effect on the reduction
process.

\Figure{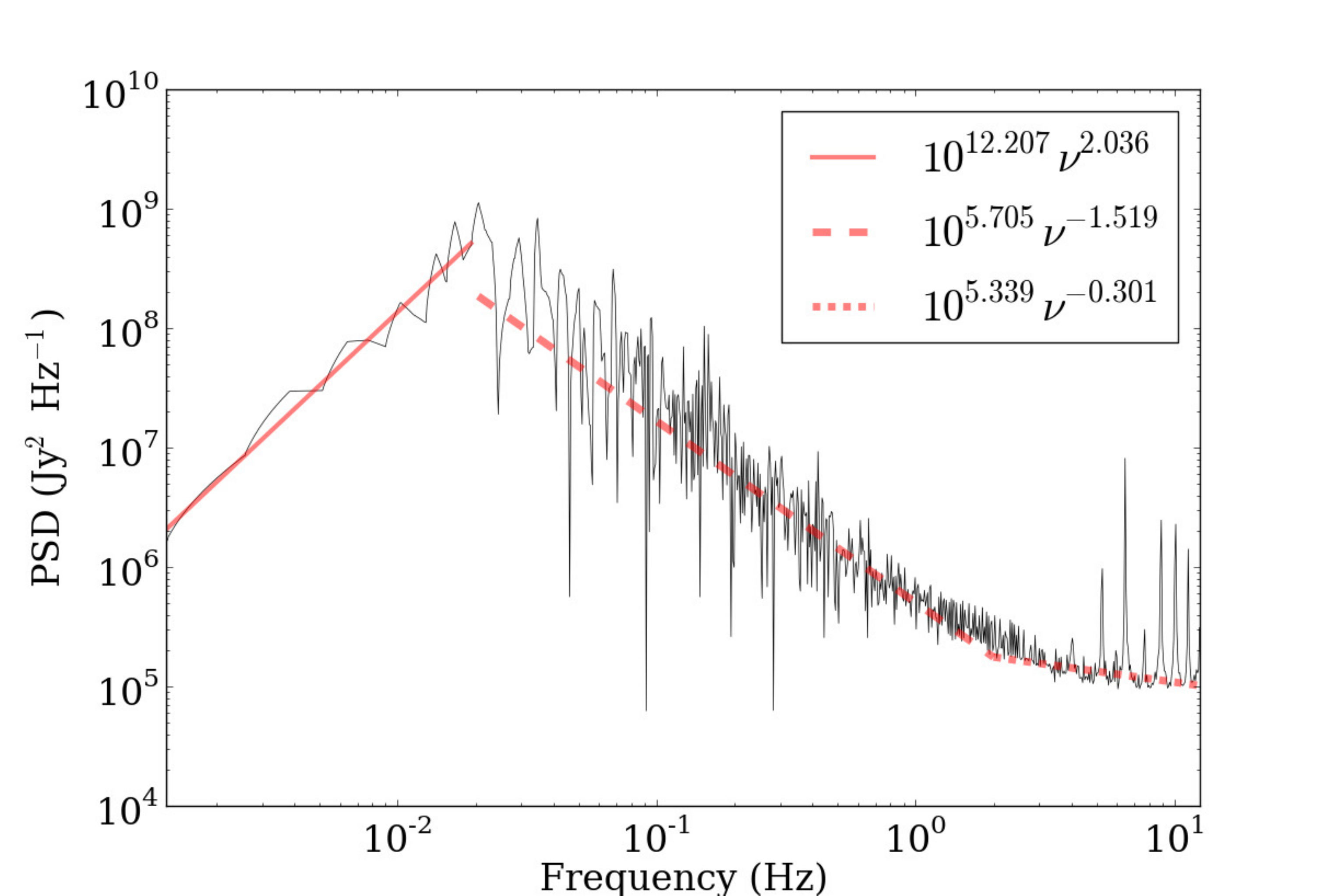}
{Fit to the raw downsampled power spectrum of a $\sim30$ minute observation.
Three independent power laws are fit to the data,  with a fixed break at 0.02
Hz (below which the AC-coupled bolometer bias and readout electronics remove
signal) and a fitted break at
higher frequency, near 2 Hz, where the power spectrum flattens towards white noise.  
The beam FWHM is at about 4 Hz using the standard scan rate of 120\arcsec s$^{-1}$.
}
{fig:powerspecfit}{0.5}{0}

The Fourier transform of the atmosphere timestream is generated by applying noise
to the fitted power spectrum.  The power at each frequency is multiplied by a
random number sampled from a Gaussian distribution%
\footnote{We experimented with different
noise distributions that reasonably matched the data, including a lognormal distribution, and
found that the angular transfer function was highly insensitive to the noise applied to the 
atmosphere time series power spectrum.}
with width 1.2, determined
to be a reasonable match to the data, and mean 1.0.
The resulting Fourier-transformed timestream $d(t)$ is $FT(d(t)) = (r_{\nu1} P_f)^{1/2} + i
(r_{\nu2} P_f)^{1/2}$, where $r_1$ and $r_2$ are the normally distributed random
variables and $P_f$ is the fitted power-law power spectrum.
The atmosphere timestream is then created by inverse Fourier transforming 
this signal.

Gaussian noise is added to the atmospheric timestream of each bolometer
independently, which renders the correlation between timestreams imperfect.
This decorrelation is important for the PCA cleaning, which would remove all of
the atmosphere with just one nulled component if the correlation was exact.
The noise level set in the individual timestreams determines the noise level in
the output map.

\subsubsection{Simulated Map Parameters}
We simulated the astrophysical sky by
randomly sampling signal from an azimuthally symmetric 2D power-law distribution in Fourier space.  The
power distribution as a function of angular frequency is given by 
\begin{equation}
    P(1/r) \propto (1/r)^{-\alpha_{ps}}
\end{equation}
where $r$ is the angular size-scale and $\alpha_{ps}$ is the power-law spectral index for power spectra. We
modeled this signal using power spectrum power-law indices ranging from -3 to
+0.5; in the HiGal $\ell=30\arcdeg$ Science Demonstration Phase (SDP) field, the power-law index measured from the
500 \um\ map is $\alpha_{ps}\sim2$ over the scales of interest for comparison
with Bolocam.%
\footnote{The Herschel data used were those presented in \citet{Molinari2010a},
and the measured power-law was consistent in more recent reductions
\citep{Traficante2011a}.}
The data were smoothed with a model of the instrument PSF to simulate the
telescope's aperture and illumination pattern.  
For each power-law index, three realizations of the map using different random seeds
were created.  
The signal map was then sampled into timestreams with the Bolocam array using
a standard pair of perpendicular boustrophedonic scan patterns.  
Examples of one of these realizations with identical random seeds and
different power laws are shown in Figure \ref{fig:exp10gridin}.

Simulations performed with $\alpha_{ps}=3$ yielded no recovered astrophysical
emission for normalizations in which the astrophysical sky was fainter than the
atmosphere.  Such a steep power spectrum is inconsistent with both BGPS and
other observations: as noted above, Herschel sees structure with
$\alpha_{ps}\sim2$. The fact that the  BGPS detected a great deal of
astrophysical signal, none of which was brighter than the atmosphere, confirms
that $\alpha_{ps}=3$ is unrealistic.

\TallFigureTwo{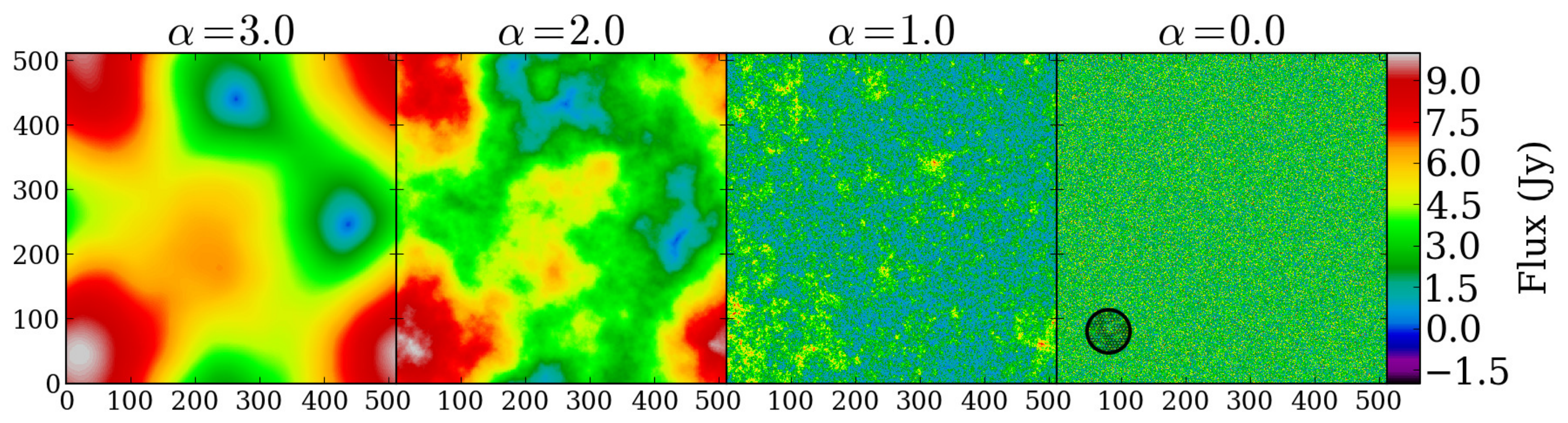}{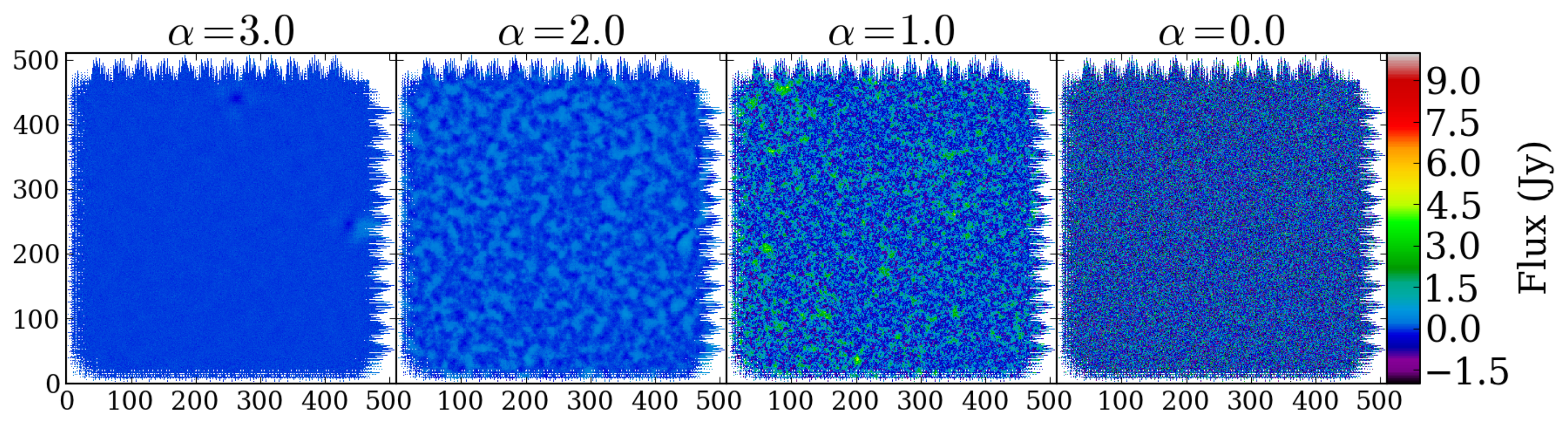}
{Examples of input (top) and output (bottom) maps for different input power-spectrum power law $\alpha_{ps}$ values.
For very steep power laws, most of the power is on the largest scales.  
$\alpha_{ps}=0$ is white noise.  The axis scales are in pixels, where each pixel
is 7.2\arcsec, so each field is approximately 1\arcdeg\ on a side.  The Bolocam
footprint is plotted with a large circle of diameter 480\arcsec and smaller
circles of diameter 33\arcsec representing each beam in its appropriate
relative location.  It is shown in the right panel of the top figure as an
indication of the largest possible recovered angular scales; it is about 1/8th
the width of the map.  The input images are normalized to have the same
\emph{peak} flux density.  The pipeline recovers no emission from the
simulation with $\alpha_{ps}=3$, but this value of $\alpha_{ps}$ is not
representative of the real astrophysical sky. 
}
{fig:exp10gridin}{1}{7in}

\subsection{The Angular Transfer Function}
\label{sec:STF}

We used a subset of these power-law simulations to measure the amount of recovered signal at each
angular (spatial) scale.
For each power-law in the range $1<\alpha_{ps}<2$, we used three different
realizations of the map to measure the angular transfer function, defined as
$STF(f) = F_{out}(f)/F_{in}(f)$ where $f$ is the angular frequency, $F_{out}$
is the azimuthally averaged power-spectrum of the pipeline-processed map, and
$F_{in}$ is the azimuthally averaged power-spectrum of the simulated input map.

The angular transfer function shows only weak dependence on the ratio of
astrophysical to atmospheric power, and is approximately constant at $\sim95\%$
recovery over the range of angular scales between the beam size and
$\sim1.5\arcmin$.  The angular transfer function is shown in Figure
\ref{fig:stf}.  At larger angular scales, in the range $2\arcmin-8\arcmin$, the
recovery is generally low ($<80\%$).  Our simulations included the full
range of observed astrophysical to atmospheric flux density ratios, from $\sim10^{-2}$
for the Central Molecular Zone (CMZ) down to $\sim10^{-4}$ for sparsely
populated regions in the $\ell\sim70\arcdeg$ region.

\Figure{{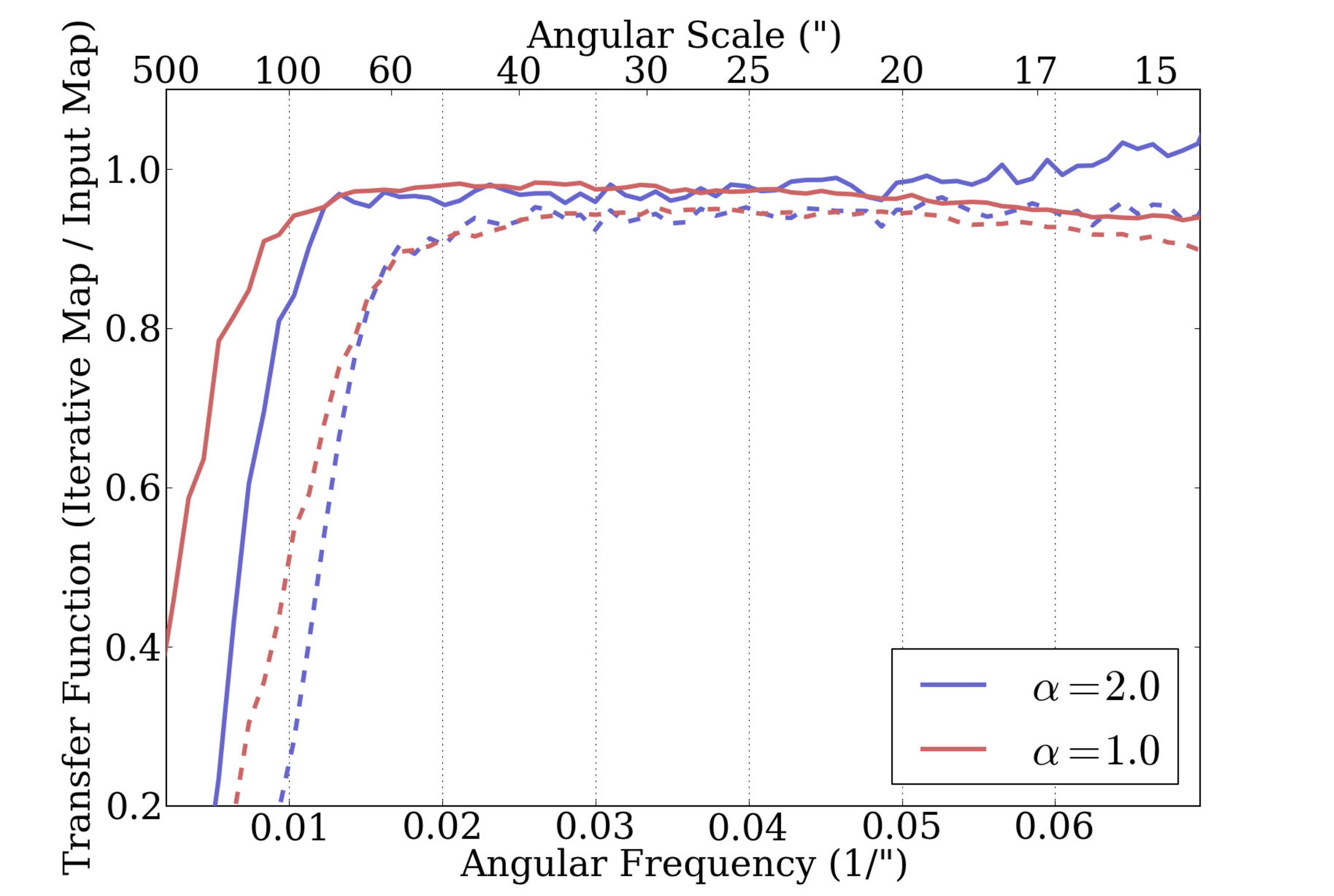}}
{The angular transfer function over the range of angular scales where the BGPS
data are reliable after 20 iterations (solid) and without iterative mapping (dashed).
At higher angular 
frequency (smaller angular scale), the beam smooths out any signal.  At lower
angular frequency, the atmospheric subtraction removes signal. 
The benefits of iterative mapping in recovered flux density on all scales, but
particularly the improvement in large-scale recovery, are evident.  The
simulations used for this measurement had a power-law sky structure with
$\alpha_{ps}=2$ (blue) and $\alpha_{ps}=1$ (red).
}
{fig:stf}{0.5}{0}

\citet{Chapin2013a} perform a similar analysis for the SCUBA-2 pipeline.  Our
transfer function (Figure \ref{fig:stf}) cuts off at a scale $\sim 1/6$ the
SCUBA-2 scale.  While the angular extent of the Bolocam
footprint is only slightly smaller than SCUBA-2's, some feature of the instrument
or pipeline allows SCUBA-2 to recover larger angular scales.  We speculate that
the much larger number of bolometers in the SCUBA-2 array allows the atmosphere
to be more reliably separated from astrophysical and internal electrical
signals (bias and readout noise), so the SCUBA-2 pipeline is able to run with
an atmosphere subtraction algorithm less aggressive than the 13-PCA approach we
adopted.

\subsection{Comparison to other data sets: Aperture Photometry}
\label{sec:otherdata}
Given an understanding of the angular transfer function, it is possible to
compare the BGPS to other surveys, e.g.  Hi-Gal, ATLASGAL, and when it is
complete, the JCMT Galactic Plane Survey (JPS), for temperature and $\beta$%
\footnote{$\beta$ is the dust emissivity index, i.e., a modification to a
blackbody to create a greybody such that $G(\nu) = B(\nu) (1-e^{-\tau(\nu)})$,
and $\tau(\nu) = (\nu/\nu_0)^\beta$.}
measurements.

Because of the severe degeneracies in temperature/$\beta$ derivation from
dust SEDs \citep[e.g.][]{Shetty2009a,Shetty2009b,Kelly2012a}, we
recommend a conservative approach when comparing BGPS data with other data
sets.  For compact sources, aperture extraction with background subtraction in
both the BGPS and other data set should be effective.  Section
\okinfinal{\ref{sec:apextract}} discusses aperture extraction in the presence of typical
power-law distributed backgrounds.

\subsection{Comparison to other data sets: Fourier-space treatment}
\label{sec:otherdata_fourier}
In order to compare extended structures, which includes any sources larger than
the beam, a different approach is required.  The safest approach is to
``unsharp mask'' (high-pass-filter) both the BGPS and the other data set with a
Gaussian kernel with FWHM $\lesssim120\arcsec$ ($\sigma \lesssim 51 \arcsec$).
The filtering will limit the angular dynamic range, but will provide accurate
results over the angular scales sampled.

Direct comparison of power spectra over the reproduced range is also possible.
A demonstration of this approach is given in Figure \ref{fig:higalpowerlaw},
which shows the structure-rich $\ell=30\degrees$ field.
The BGPS power spectrum has a shape very similar to that of HiGal. 
The spectral index is a commonly used measure of the ratio between
flux densities at two different wavelengths in the radio, 
\begin{equation}
    \frac{F_2}{F_1} = \left(\frac{\lambda_2}{\lambda_1}\right)^{-\alpha_{\nu}} = \left(\frac{\nu_2}{\nu_1}\right)^{\alpha_{\nu}}
\end{equation}
The spectral index between the BGPS at 1.1 mm and Herschel at 500 \um is
$\alpha_{\nu}\sim 3.7$ over the range $33\arcsec < dx < 300\arcsec$, although because the
BGPS angular transfer function is low at the large end of this range, this is
only an `eyeball' estimate.  On the Rayleigh-Jeans tail,
$\alpha_{\nu}=\beta+2$, so this spectral index is consistent with typical dust
emissivity index $\beta$ measurements in the range $1.5 < \beta < 2$. 

\Figure{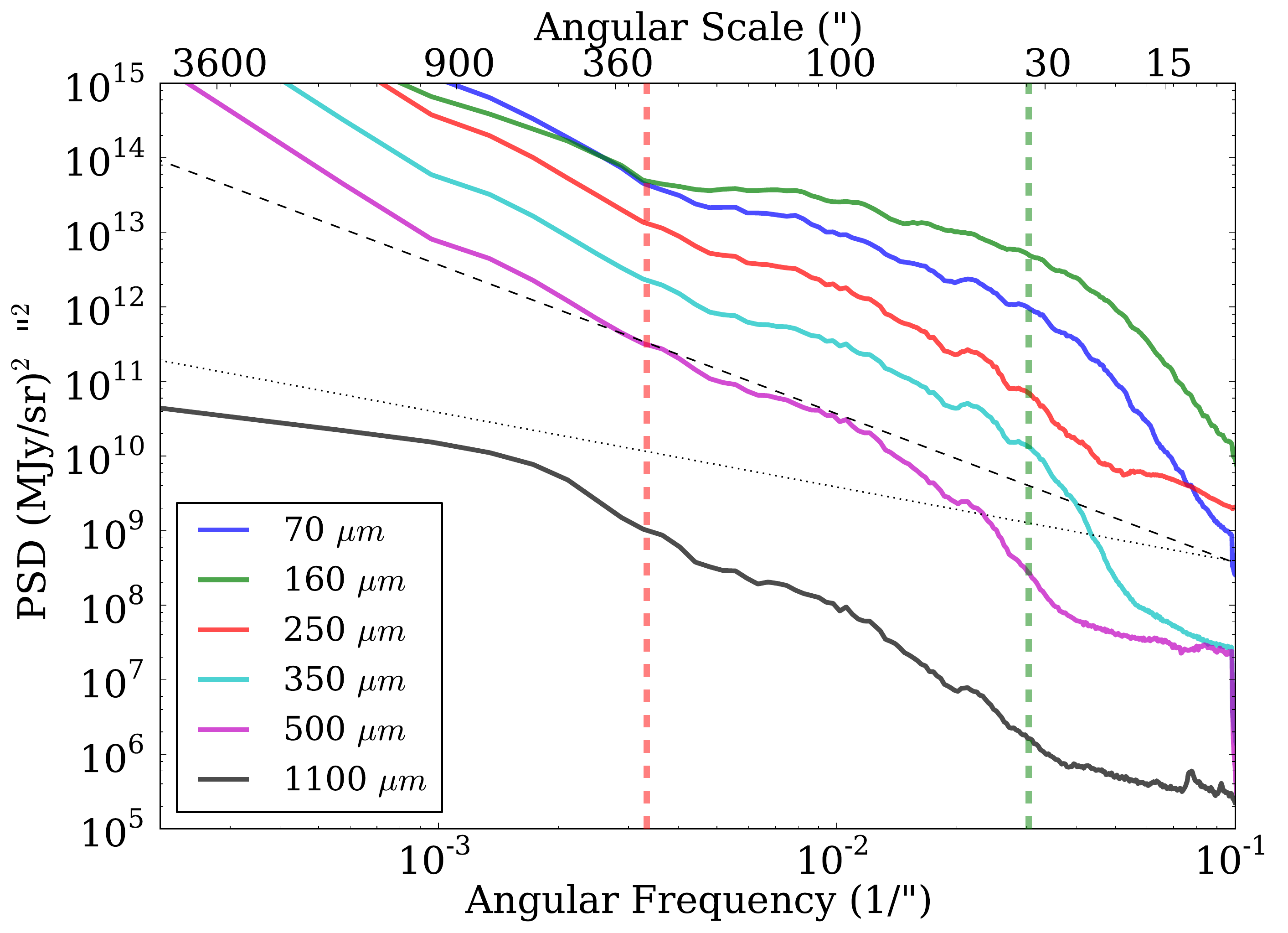}
{A comparison of the power spectra of the $\ell=30\arcdeg$ HiGal SDP fields
with the BGPS power spectrum covering the same area.  The area included is 1
square degree.  The dashed and dotted black lines indicate power laws with
$\alpha_{ps}=2$ and $\alpha_{ps}=1$ respectively, with arbitrary
normalizations, as a guide for comparison.  The vertical dashed red and green
lines indicate the large angular scale 50\% recovery point of the BGPS (given
an $\alpha_{ps}=1$ input) and the BGPS beam FWHM respectively.  The ratio of
500 \um to 1100 \um in this example and over the marked range has a spectral
index $\alpha_{\nu}\sim3.7$.  Note that the 500 \um power begins falling off
more steeply at $\sim40\arcsec$ because the Herschel FWHM beam size is
$\sim42\arcsec$ at 500 \um, slightly larger than Bolocam's \citep[at 250 and
350 \um, the Herschel beam is $\sim23\arcsec$ and $30\arcsec$,
respectively;][]{Traficante2011a}.}
{fig:higalpowerlaw}{0.6}{0}

\section{Source Extraction}
\label{sec:sourceextraction}
\citet{Rosolowsky2010} presented the Bolocat catalog of sources extracted from
the \vone data with a watershed decomposition algorithm.  We have used the same
algorithm to create a catalog from the \vtwo catalog.  We have also performed
comparisons between the \vone and \vtwo data based on the extracted sources.
The new catalog was derived 
using the same Bolocat parameters as in \vone.  This catalog
includes regions that were not part of the \vone survey area, but we restrict
our comparison between \vone and \vtwo to the area covered by both surveys.  

\subsection{Aperture Extraction}
\label{sec:apextract}

One major change from the \vone catalog is that the fluxes in the \vtwo
catalog are reported with background subtracted.  The backgrounds are calculated
from the mode of the pixels in the range $[2R,4R]$, where R represents the aperture
radius (20\arcsec, 40\arcsec, or 60\arcsec).  The mode is computed using 
the IDL astrolib routine \texttt{skymod.pro}, which returns the mean of the
selected data if the mean $\bar{\mu}$ is less than the median $\mu_{1/2}$
(indicating low ``contamination'' from source flux) or $3 \mu_{1/2} - 2
\bar{\mu} $ otherwise, then performs iterative rejection of bad pixels
\citep{Landsman1995a,Stetson1987a}.  

We performed aperture extraction on simulated data sets to determine what size
apertures are appropriate when comparing to other data sets.
In Figure \ref{fig:bgsub}, we show the results of aperture extraction with and
without background subtraction on a simulated power-law generated image with
$\alpha_{ps}=2$ before and after pipeline processing.  The map has had point
sources added to it randomly distributed throughout the field with flux
densities randomly sampled from the range $[0.1,1]$ Jy, and the power-law
extended flux has an amplitude $\approx1.8$ Jy.  Sources are extracted from the
pipeline-processed map using Bolocat, then the same source locations and masks
are used to extract flux measurements from the input map.  Figure \ref{fig:simptsrc} shows
the input, pipeline-processed, and point-source-only maps along with the Bolocat apertures to give
the reader a visual reference for an $\alpha_{ps}=2$ background with point sources.
The scatter between
the flux density measurements derived from the input simulated sky map and the
iteratively produced map is small when background subtraction is used (the blue
points), but large and unpredictable otherwise (the red points).

The agreement between the flux densities extracted from the iterative map and
the input synthetic map is excellent for 40\arcsec\ diameter
background-subtracted apertures.  For these apertures, the RMS of the
difference between the iterative map and the input map fluxes is $\sigma=0.03$
Jy when background subtraction is used, indicating the utility and necessity of
this approach.  The agreement is similarly good for 80\arcsec apertures
($\sigma=0.10$ Jy), but the 120\arcsec apertures exhibit a source- and
background-brightness dependent bias, so we recommend against apertures that
large when comparing to other data sets.

\FigureTwo{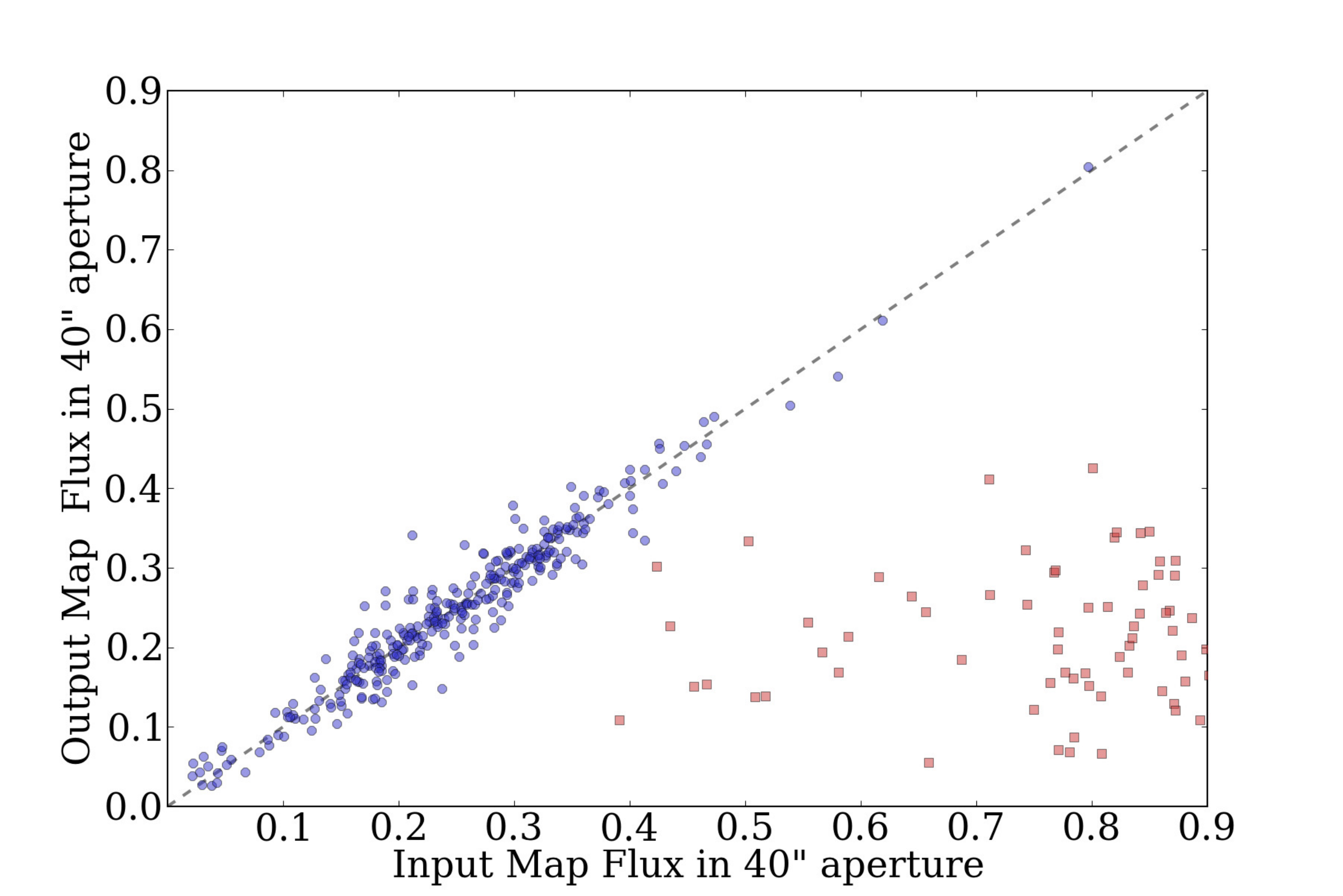}{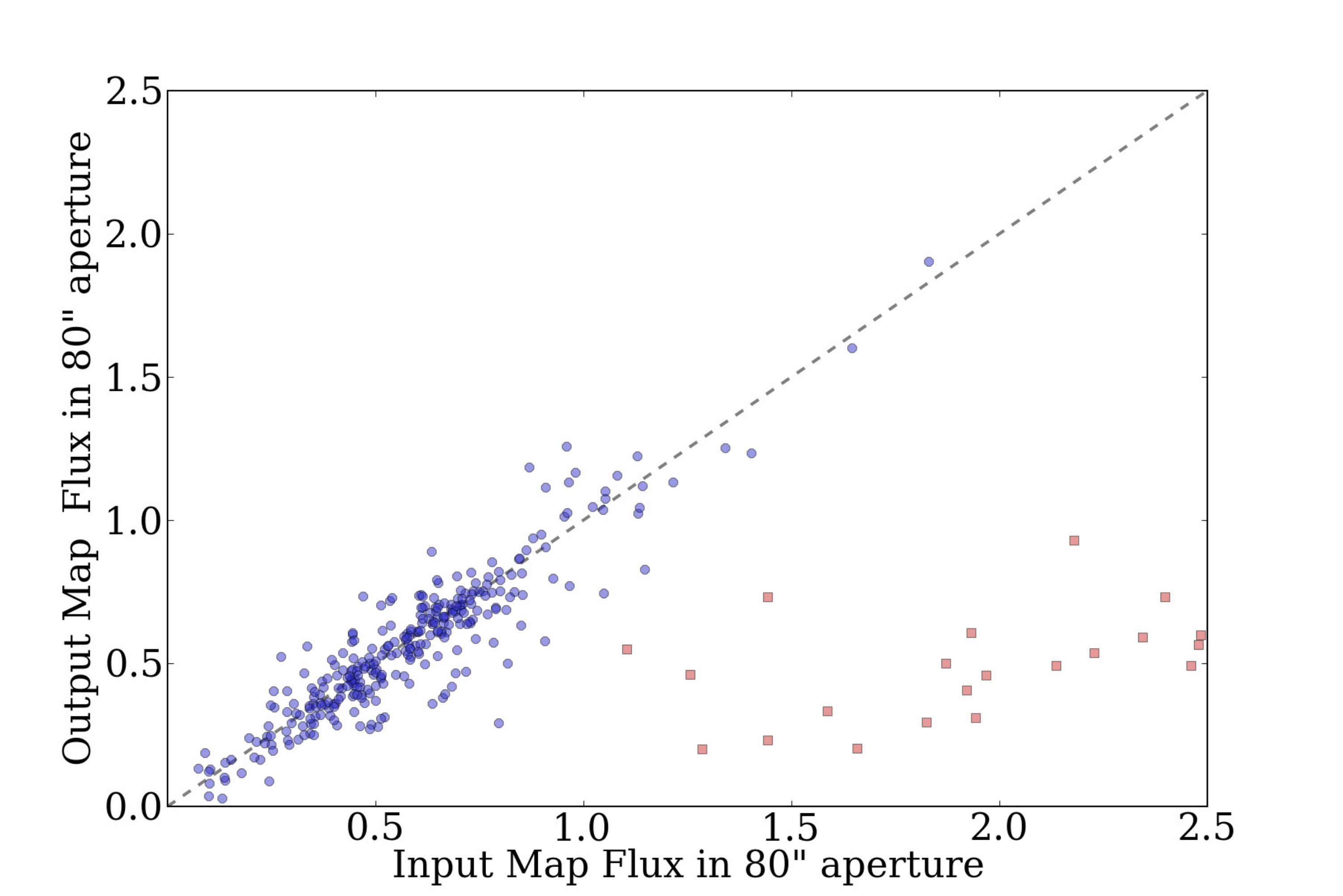}
{The aperture-extracted flux densities in a simulated map.  Sources are
identified from the pipeline-processed map, then flux densities are extracted
from both the unprocessed input map and the pipeline-processed map.  The X-axis
shows the flux density of the source in the input map with (blue circles) and
without (red squares) the flux density in a background annulus subtracted.
Many of the red sources are not displayed as they are far to the right side of
the plot, indicating poor agreement between the input and processed maps.  The
Y-axis shows the flux density extracted in the same aperture from the output
pipeline-processed map.  The black dashed line is the 1-1 line.  The left plot
shows 40\arcsec and the right plot 80\arcsec diameter apertures.    Section
\ref{sec:sourceextraction} describes the background subtraction process; the
\vtwo catalog reports background-subtracted flux density measurements.  }
{fig:bgsub}{1.2}

\Figure{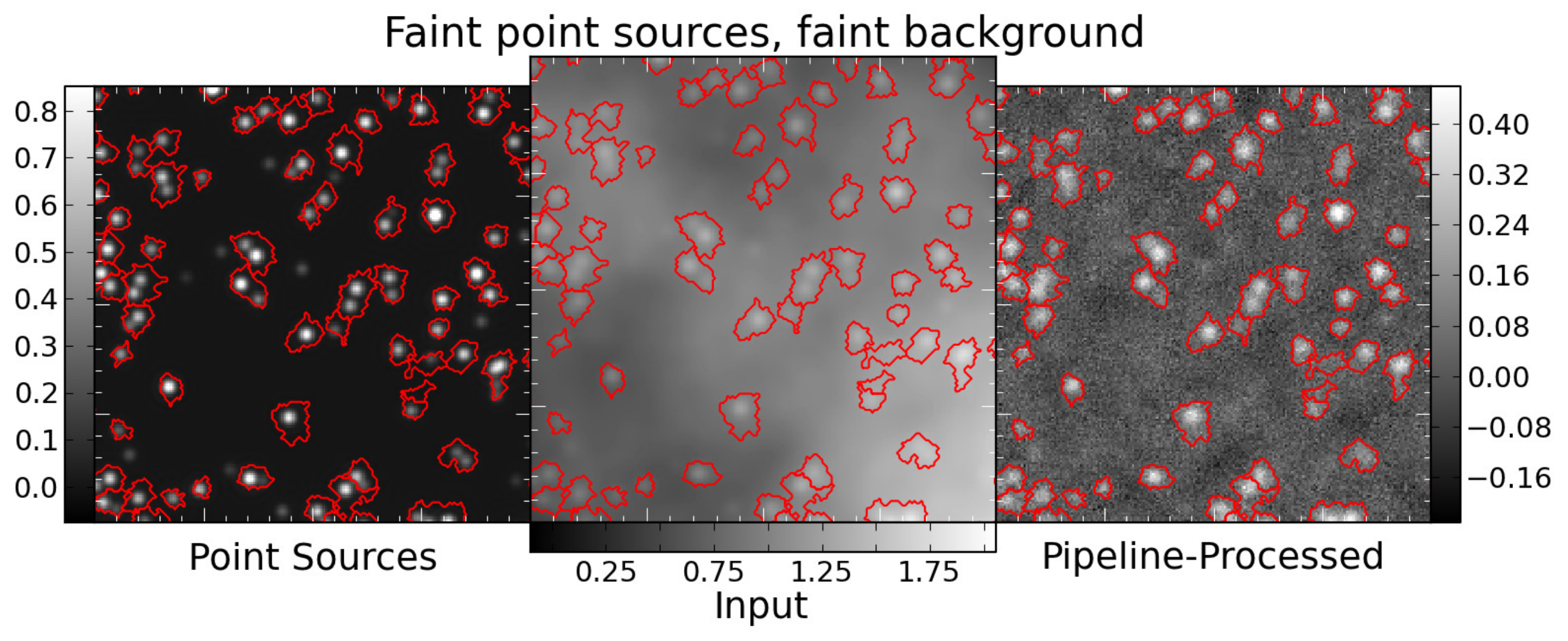}
{Images from a simulation of a power-law distributed background with
$\alpha_{ps}=2$ and point sources with peak flux densities in the range
$[0.1,1]$ Jy/beam.  The left panel shows the pipeline-processed map, which was
used to define the Bolocat masks shown as red contours.  The colorbars show the
flux density in units of Jy/beam.  The power-law flux density distribution is
evident as the structure between point sources in the left image; it is only
weakly recovered by the pipeline because most of the power is on large angular
scales and therefore filtered out.
}{fig:simptsrc}{0.6}{0}

There are caveats to this analysis.  If the ``background'' power-law map has
a peak flux density $\gtrsim10\times$ the peak point-source flux density, the point
sources will not be recovered: the cataloging algorithm will pick out peaks in
the power law flux distribution.  These cannot be analyzed with simple aperture
extraction for an $\alpha_{ps}=2$ flux density distribution.  However, for
shallower power-law distributions, i.e. $\alpha_{ps}\lesssim1$, aperture
extraction effectively recovered accurate flux-densities in the processed maps
- shallow power-law distributions more strongly resemble point-source-filled
maps.

\subsection{Catalog Matching between \vone and \vtwo}
\label{sec:catalogmatching}
We matched the \vone\ and \vtwo\ catalogs based on source proximity. 
For each source in \vone, we identified the nearest neighbor from \vtwo, and found
that 5741 \vtwo sources are the nearest neighbor for a \vone source out of 8004 \vtwo sources
in the \vone-\vtwo overlap region.  Similarly,
we identified the nearest neighbor in \vone for each \vtwo source, finding 5745
\vone sources are the nearest neighbor for a \vtwo source out of 8358 \vone sources.  There are 5538
\vone-\vtwo source pairs for which each member of the pair has the other as its
nearest-neighbor.  These sources are clearly reliable and stable source identifications and constitute
about 70\% of the \vtwo sample.

Most of the unmatched sources
have low flux density (Figure \ref{fig:v1vsv2matchhist}), but some were significantly higher -
these generally represent sources that were split or merged going from \vone to \vtwo.
A few examples of how mismatches can happen are shown in Figures \ref{fig:contourmatch} and \ref{fig:contourmatch2}.
The low-flux-density sources were most commonly unmatched in regions where the noise in
\vone and \vtwo disagreed significantly.  The
high-flux-density mismatches tend to be different decompositions of bright sources and
are preferentially found near very bright objects, e.g. in the Galactic center
region.

\Figure{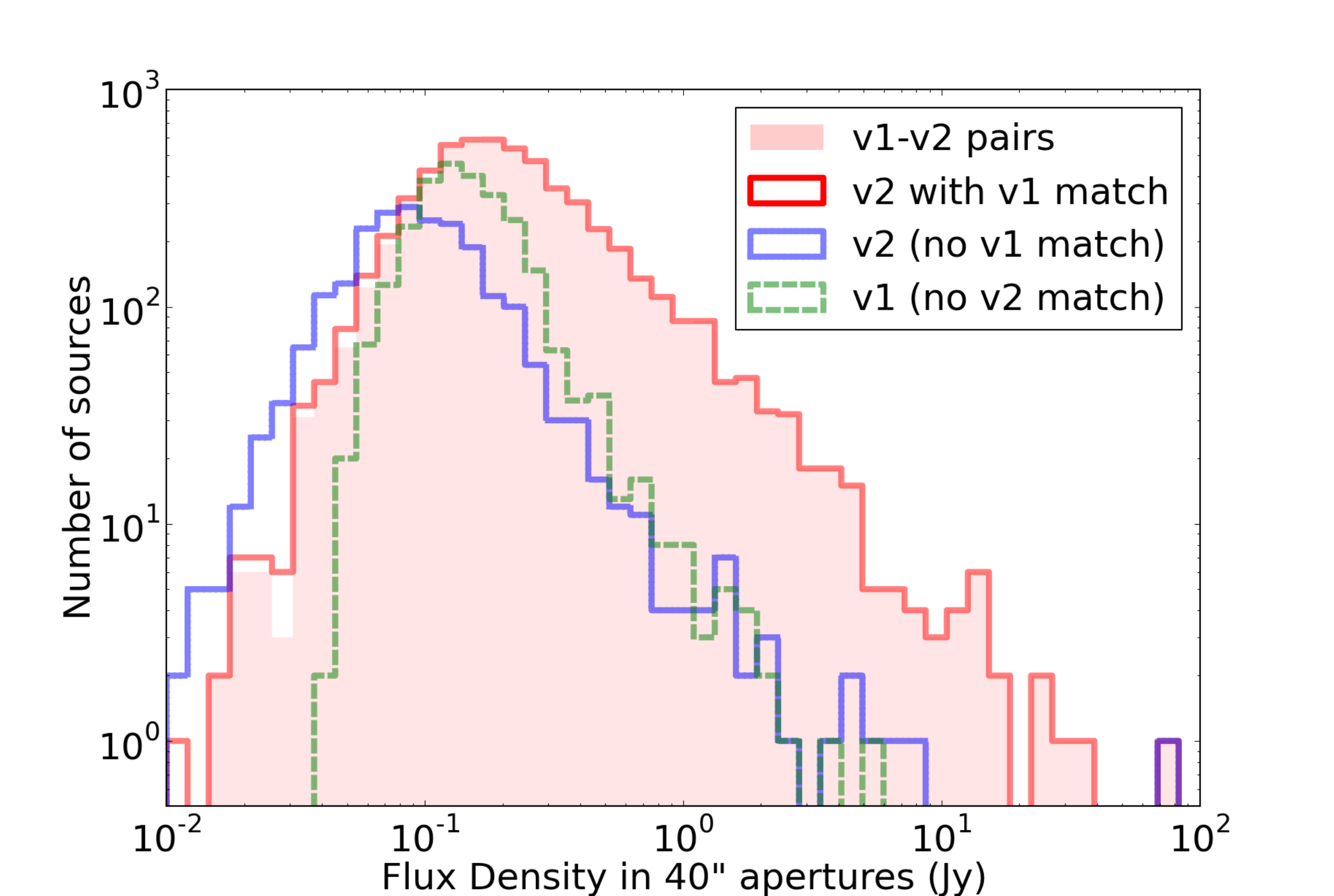}
{Histograms showing the sources matched between the \vone\ and \vtwo\ catalogs.
Most of the \vtwo sources (5741 of 8004 \vtwo sources in the \vone-\vtwo overlap
region) have matches from \vone, but there is a
substantial population with no match.  The unmatched sources tend to have
lower flux densities.  The shaded area shows 1-1 matches, while the solid red
line shows one-way (unreciprocated) matches.
}
{fig:v1vsv2matchhist}{0.5}{0}

\Figure{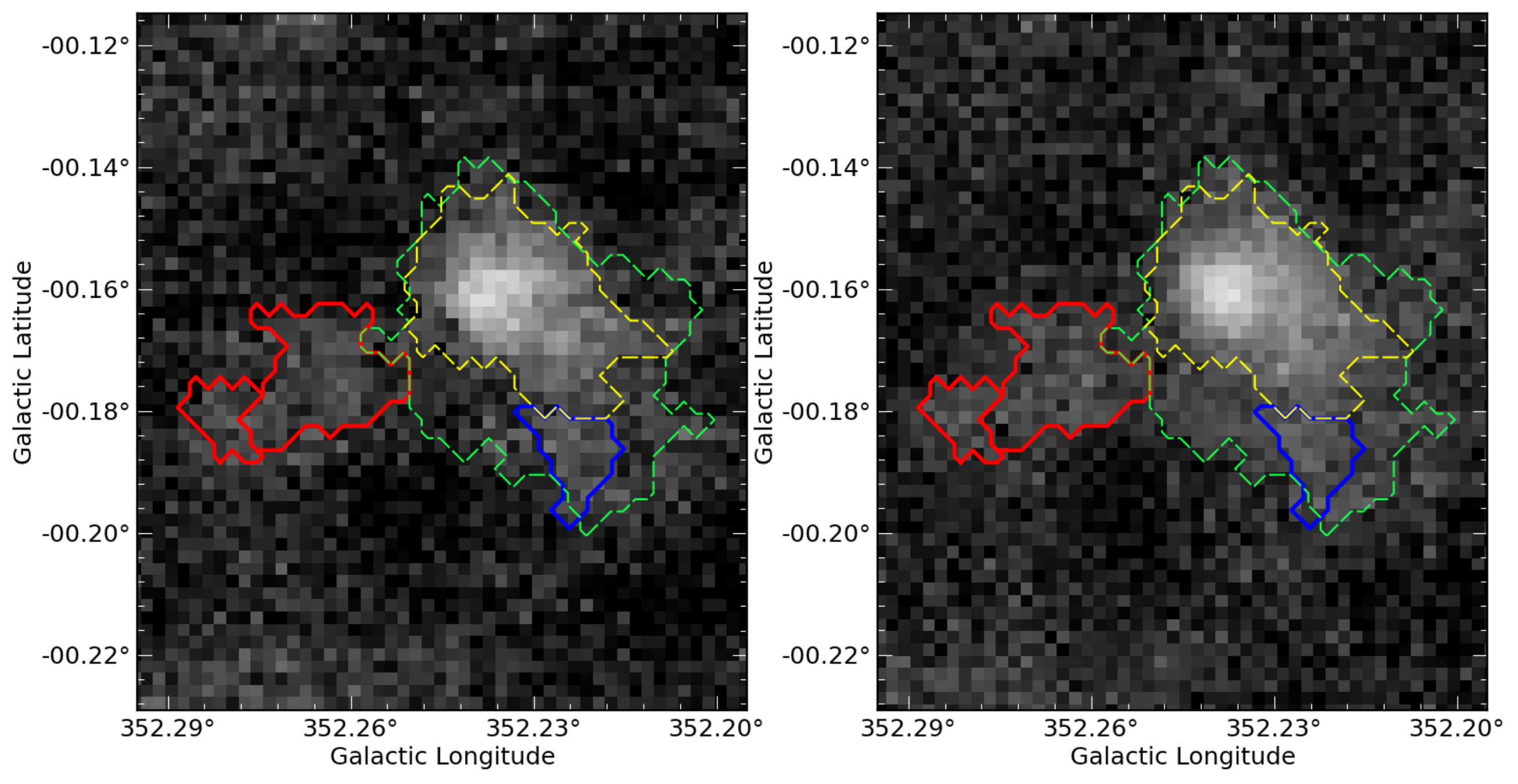}
{Contours of the extracted sources overlaid on grayscale images of a region in
\vone (left) and \vtwo (right).  The \vone data are scaled up by the
$1.5\times$ calibration correction.  The red contours show new \vtwo sources
with no \vone match, while the blue contours show \vone sources with no \vtwo
match.  The green and yellow contours show \vtwo and \vone sources with a
one-to-one match, respectively.  In this example, the \vtwo source is
significantly larger than the \vone source and merges with a shoulder that was
classified as a separate source in \vone.  Additional \vtwo sources are
detected because of increased signal-to-noise in
the red-contoured regions.}
{fig:contourmatch}{0.6}{0}

\Figure{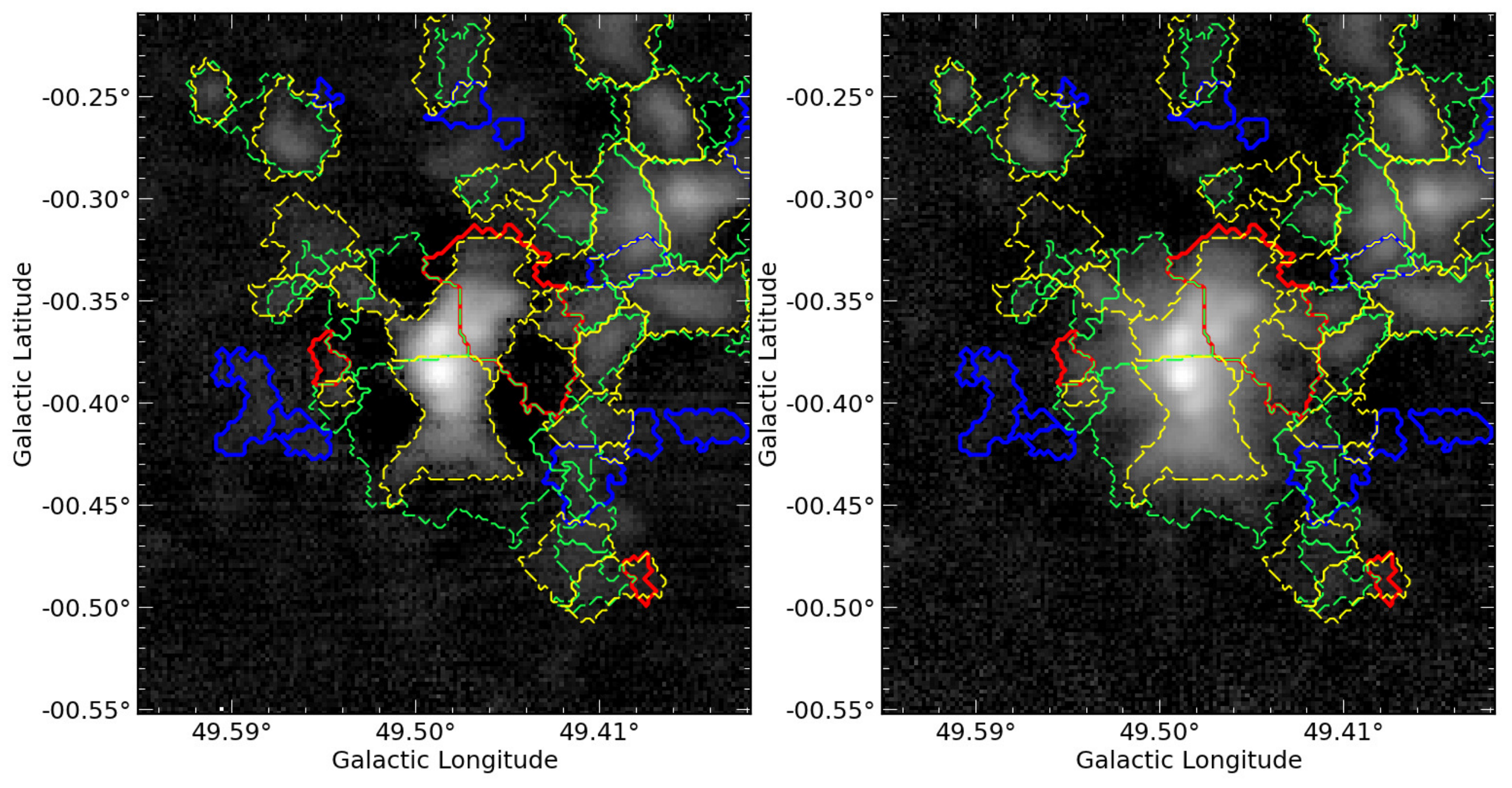}
{Same as Figure \ref{fig:contourmatch}, but for the W51 complex. The area
displayed is larger in order to encompass the entire source structure.  The
\vtwo sources are larger than the corresponding \vone sources because the
negative bowl structures have been filled in.  The red contours show regions
where \vtwo sources were detected, but because of crowding no nearest-neighbor
pair was identified in \vone: there are more \vtwo sources than \vone sources.
In this region, the brightest \vtwo sources are larger and brighter, but there
are fewer fainter sources than in \vone.}
{fig:contourmatch2}{0.6}{0}

\subsection{Source flux density, size, shape, and location distributions}
We reproduce parts of \citet{Rosolowsky2010} Figures 17 and 19 as our own
Figures \ref{fig:bolocathistograms} and \ref{fig:sizehistograms}.  These
figures show the distributions of extracted source properties (flux density, size, and
aspect ratio) for the \vone\ and \vtwo\ data.  The source flux density distributions
above the completeness cutoff are consistent between \vone and \vtwo, both exhibiting
power-law flux density distributions 
\begin{equation}
    \frac{dN}{dS_{\nu}} \propto S_{\nu}^{-\alpha_{src}}
\end{equation}
with values in the range $\alpha_{src}=2.3-2.5$ for sources with
$S_\nu\gtrsim0.5$ Jy.  In the left panel of Figure \ref{fig:bolocathistograms},
we have included the \vtwo aperture-extracted data both with and without
annular background subtraction.  The \vone catalog had no background
subtraction performed because the backgrounds were thought to be negligible,
but the \vtwo catalog has had background subtraction performed so that the flux
densities reported more accurately represent the sky.  The \vtwo data include
more large sources.  

The longitude and latitude source flux density distribution plots, Figure 15 of
\citet{Rosolowsky2010}, are reproduced in Figure \ref{fig:latlonhist}.  The properties are
generally well-matched, although even with the 1.5$\times$ correction factor to
the \vone\ data, there is more flux density per square degree in \vtwo\
sources.  The gain in flux density recovery is both because of an increased
flux density recovery on large angular scales and because of improved noise
estimation, which results in a greater number of pixels being included in
sources (see Section \ref{sec:catalogmatching} for more details and Figures
\ref{fig:contourmatch} and \ref{fig:contourmatch2} for examples).

A two-dimensional histogram providing a broad overview of the survey contents
is shown in Figure \ref{fig:hist2d}.  The ratio of source counts per half
square degree is included in panel 3.  This figure illustrates that the two
catalog versions are broadly consistent, and the regions in which they differ
significantly tend to have fewer sources.  The most extreme ratios of \vtwo to
\vone source counts tend to occur along field edges both because of
preferentially low source counts and because the \vtwo images have slightly
greater extents in latitude than the \vone.

\FigureTwo{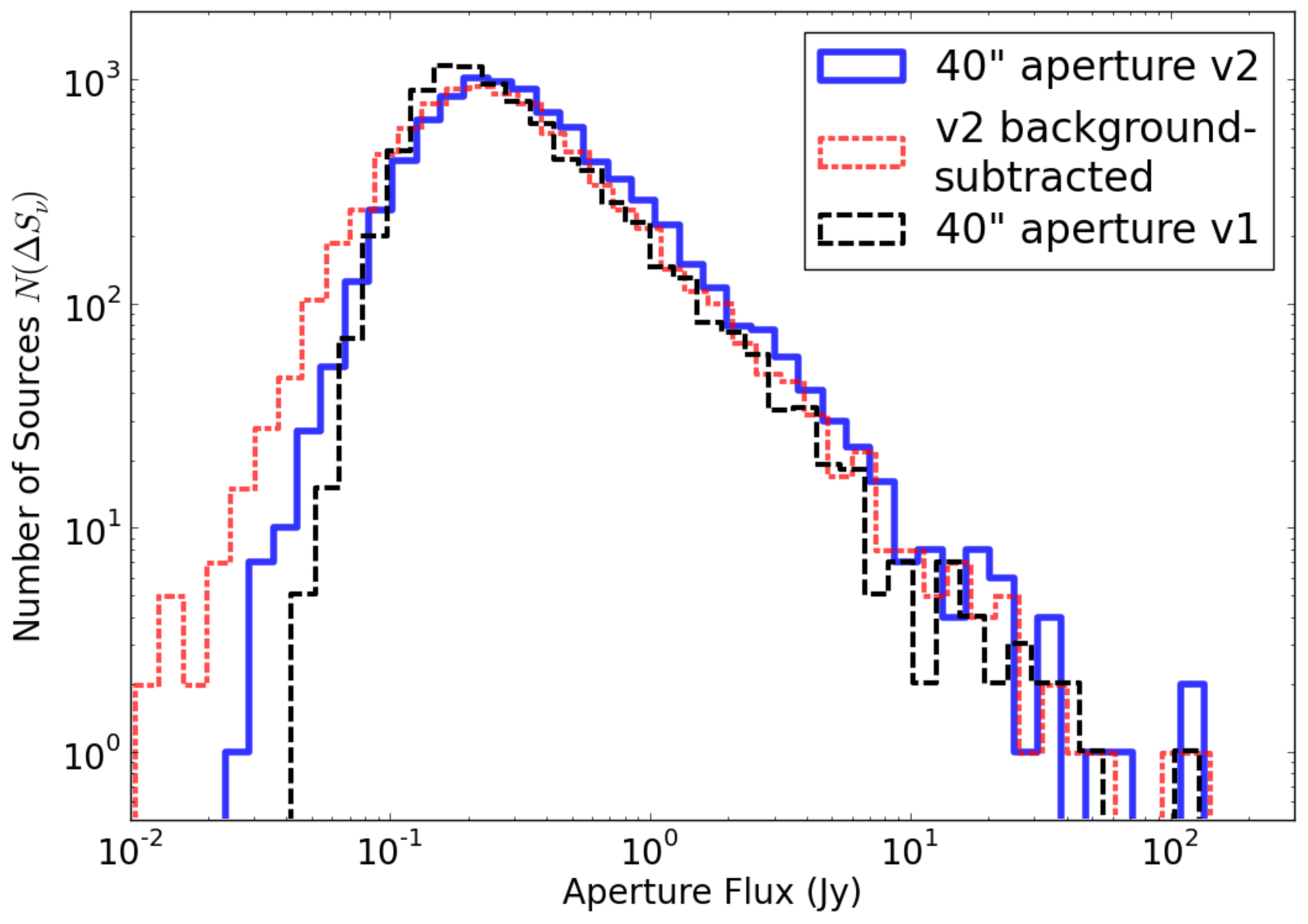}
{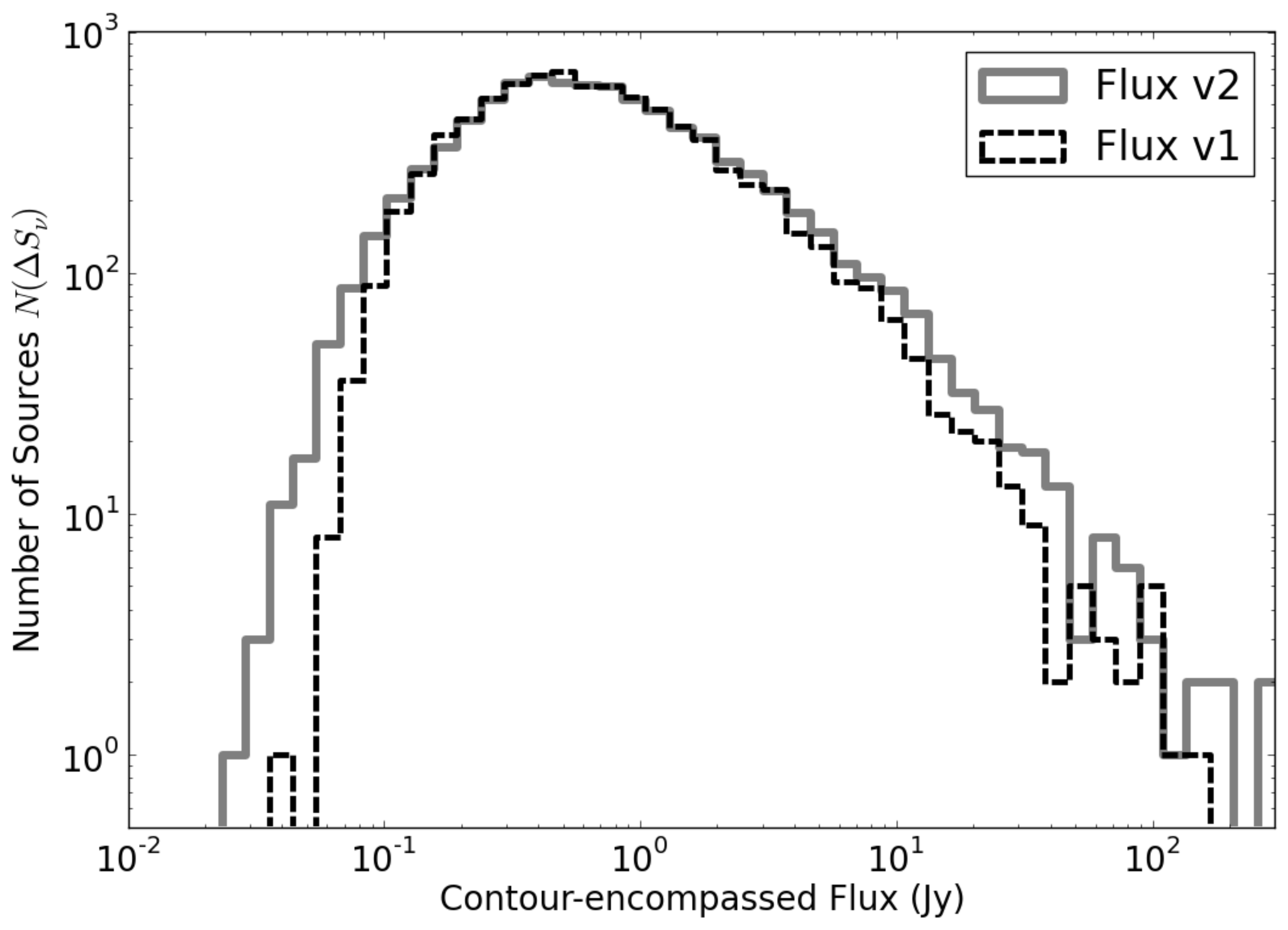} {
Comparisons of \vone and \vtwo flux density histograms.  (\textit{left}) Flux density
distribution within 40\arcsec diameter apertures.  The 40\arcsec apertures show
the \vtwo data both with and without annular background subtraction; the \vone
data are not background-subtracted.  The histogram lines are slightly offset in
order to minimize overlap.
(\textit{right}) Flux density distribution in
contour-defined apertures. No background subtraction is performed for the
contour-based flux densities in either version.}
{fig:bolocathistograms}{1}

\FigureTwo{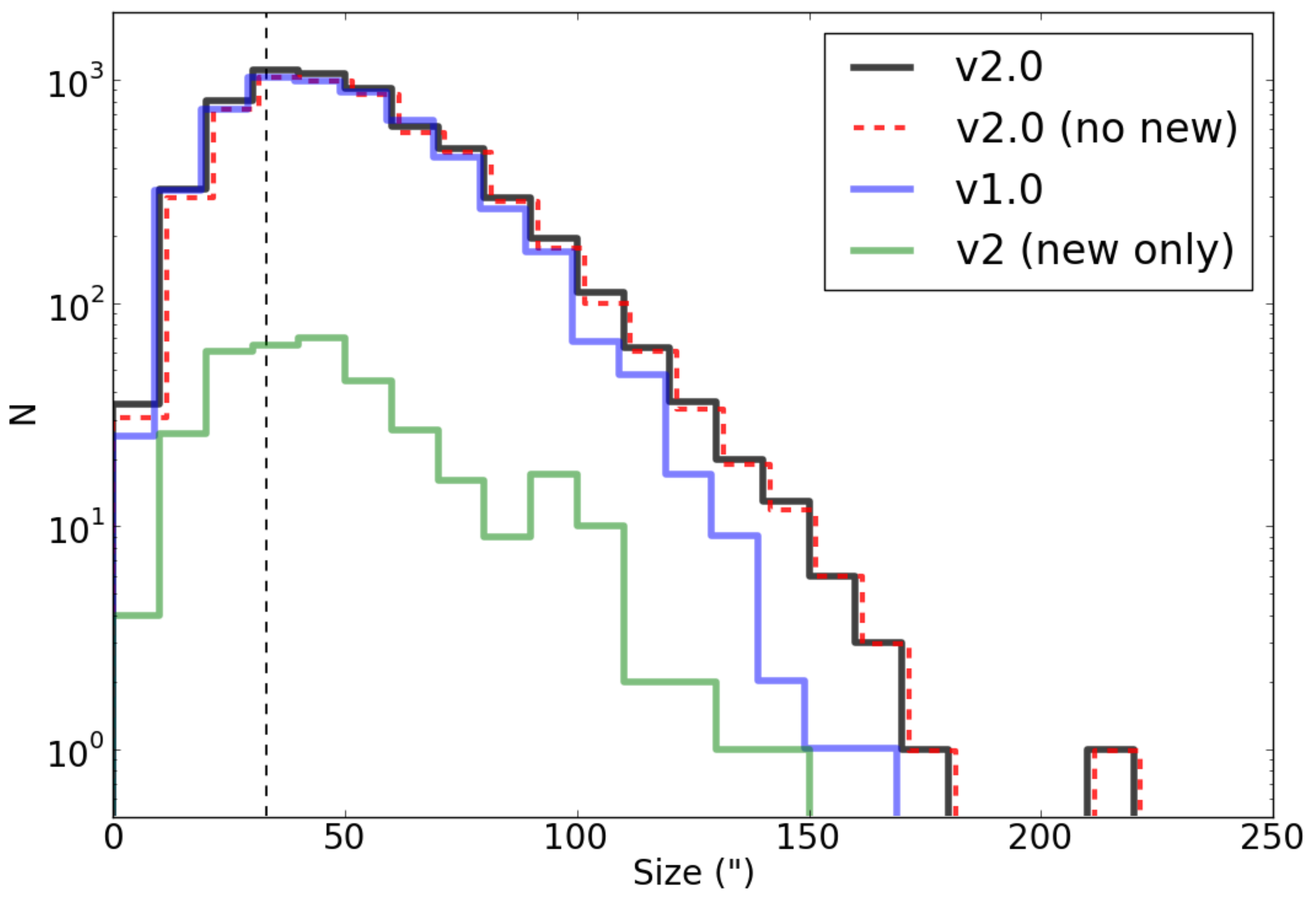}
{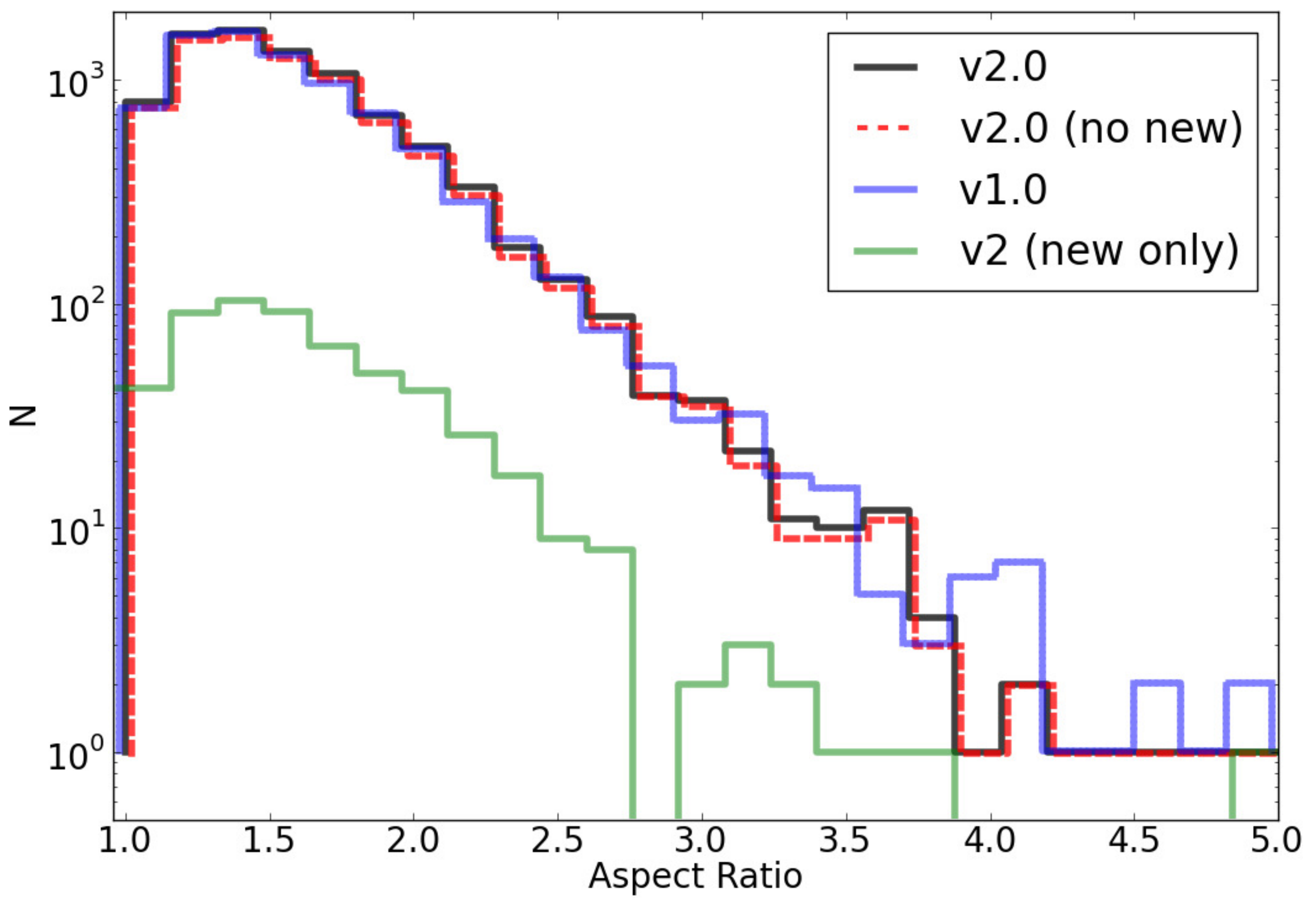}
{ Distributions of deconvolved angular sizes (left) and aspect ratios
(right) of sources in the BGPS catalog.  The vertical dashed line in the left
figure is plotted at the FWHM of the beam.  The BGPS \vtwo includes newly
observed regions not in the \vone survey, so separate histograms excluding the
new (red dashed) and excluding the old (green solid) regions are shown.  In
both plots, the histograms are slightly offset to reduce line overlap.}
{fig:sizehistograms}{1}

\FigureTwo{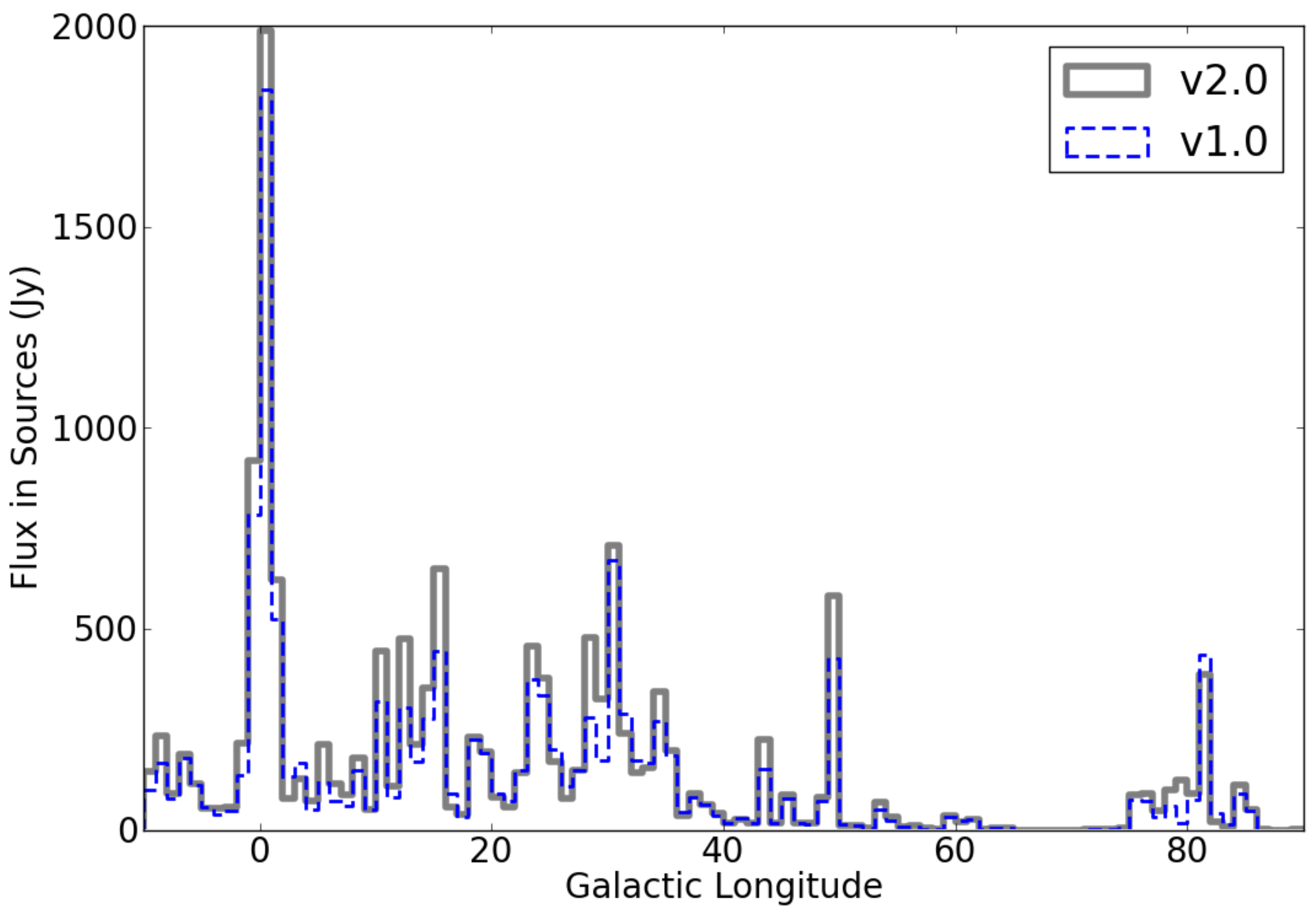}
{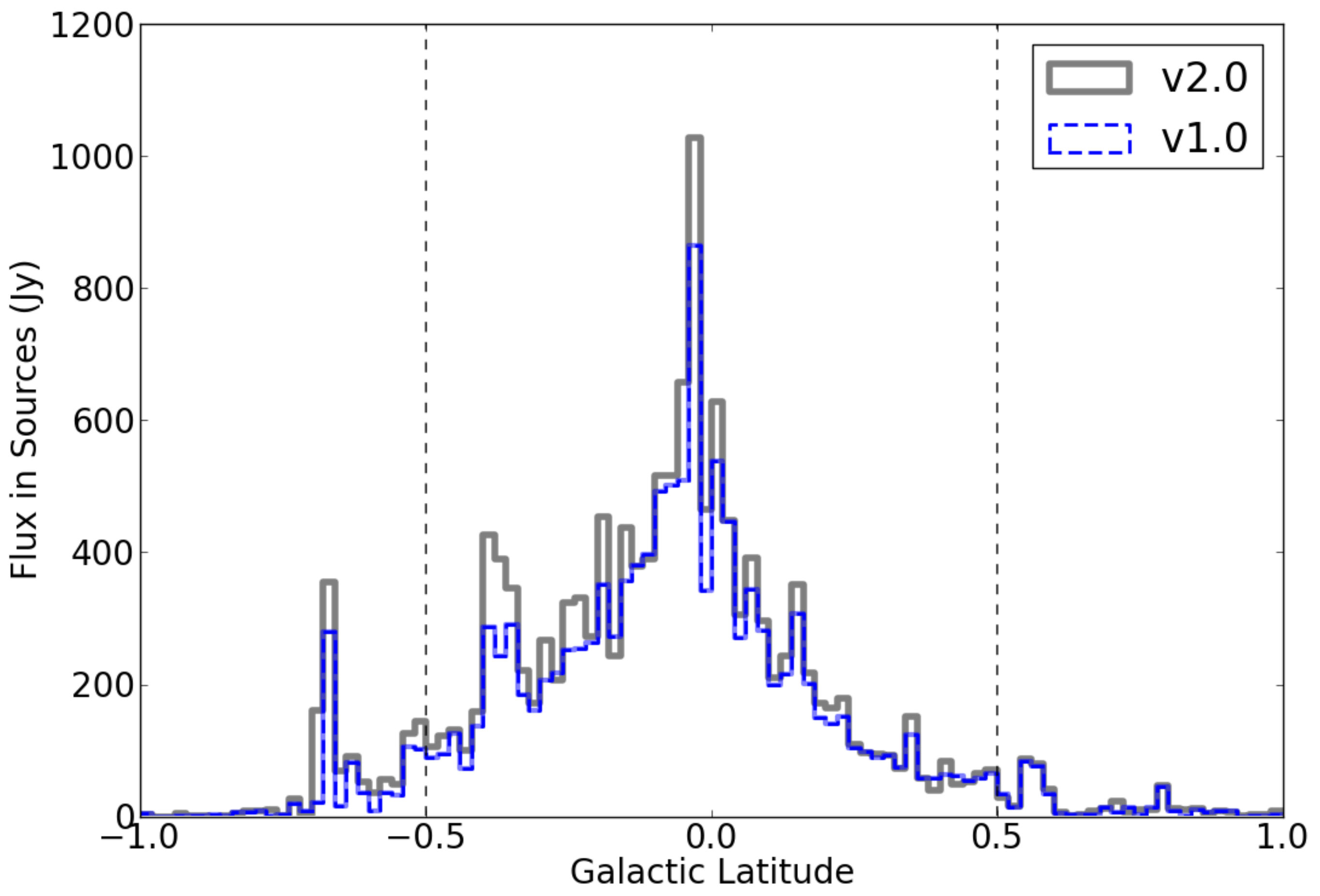}
{Distribution of total flux density in catalog sources as a function of
longitude (\textit{left}) and latitude (\textit{right}) in the Galactic plane.
The distributions contain sources extracted in the $-10 \arcdeg < \ell < 90
\arcdeg$ region.  (\textit{right}) Vertical dashed lines indicate the extent of
complete coverage in the latitude direction ($\pm0.5 \arcdeg$).  The large
excess in \vtwo compared to \vone at $b\sim-0.4$ is due to the W51 complex, in
which the flux density recovered in \vtwo was $1.5\times$ greater than in \vone,
largely because of reduced negative bowls around the brightest two sources (see Figure \ref{fig:contourmatch2}).
}{fig:latlonhist}{1}

\Figure{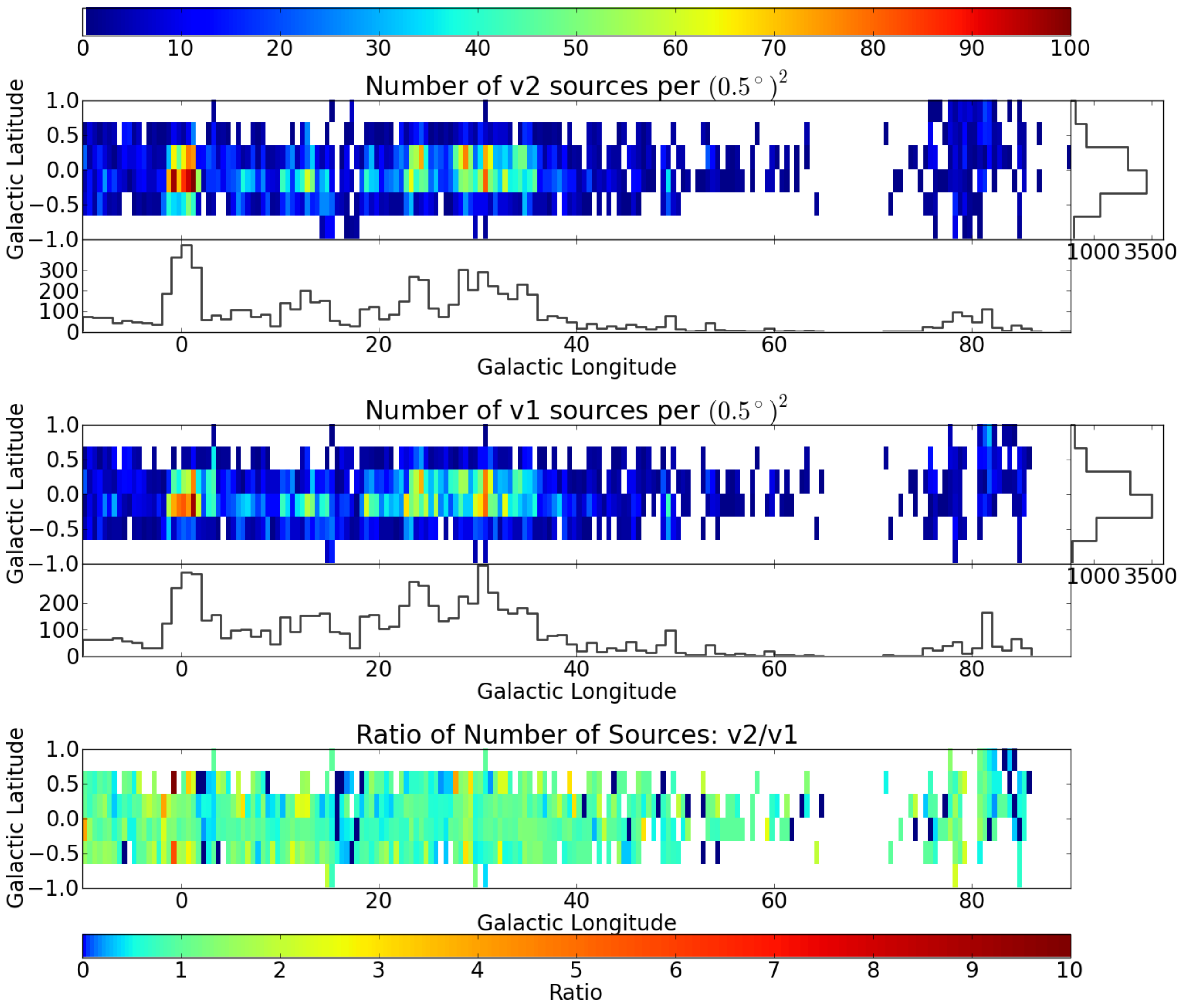}
{The two-dimensional distribution of source counts in both \vone and \vtwo.  The
colors in the first two panels illustrate the number of sources per
half-degree-squared bin as indicated by the top colorbar.  The bottom colorbar
labels the ratio of the count of \vtwo to \vone sources.  The histograms are
coarse versions of Figure \ref{fig:bolocathistograms} and show the projection
of the 2D histograms along each axis.   A preference toward negative-latitude
sources is evident at $\ell<60\arcdeg$, corresponding to our view of the Galaxy from
slightly above the plane.}
{fig:hist2d}{0.45}{0}

\section{Conclusions}
We presented Version 2 of the Bolocam Galactic Plane Survey, which is a
significant improvement over \vone in pointing and flux calibration accuracy.
The \vtwo data show an improvement in large angular scale recovery.
The \vtwo release includes new observations of regions in the outer galaxy.

\begin{itemize}
    \item We have characterized the angular transfer function of the Bolocam pipeline.
    Flux recovery is $>95\%$ for scales between $33\arcsec < dx \lesssim 80\arcsec$.
    The angular transfer function shows a sharp drop in recovered power above
    $\gtrsim100\arcsec$ scales.

    \item We compared the pointing of the BGPS to that in Hi-Gal, and found that the surveys
are consistent to within the measurement error $\sigma\approx3.5\arcsec$.

    \item We measured the power spectral density in some regions and compared
it to that in Hi-Gal, concluding that the power spectra are consistent
with the normally used dust emissivity values in the range $\beta \sim1.5-2$.

    \item A new version of the catalog has been released. The improved quality of the
\vtwo images has some effects on the BGPS catalog but the basic statistical
properties of the catalog have not significantly changed.  Because of changing
noise properties within the images, only 70\% of the individual sources in \vtwo
have an obvious \vone counterpart and vice versa.  The remaining 30\% of sources
do not have obvious counterparts because of two effects: 
\begin{enumerate}
    \item At low
significance, changing noise levels recover different features at marginal
significance.  It is likely that low significance sources in \vone and \vtwo are {\em
both} real features but have been rejected in the other catalog because of the
relatively conservative limits placed on catalog membership.  
    \item At high
significance, the catalog algorithm is dividing up complex structure using the
underlying watershed algorithm.  In this case, the precise boundaries between
objects are sensitive to the shape of the emission.  All of the high
significance features appear in both catalogs, but the objects to which a
piece of bright emission is assigned can vary.   
\end{enumerate} 
\end{itemize}
Despite these changes in the
catalogs, the overall statistical properties of the population show little
variation except that the largest sources appear brighter and larger owing to
better recovery of the large scale flux density.

\textbf{Acknowledgements:}
We thank the referee for a timely and constructive report.
This work was supported by the National Science Foundation through NSF grant
AST-1008577.  The BGPS project was supported in part by NSF grant AST-0708403,
and was performed at the Caltech Submillimeter Observatory (CSO), supported by
NSF grants AST-0540882 and AST-0838261.  The CSO was operated by Caltech under
contract from the NSF.  Support for the development of Bolocam was provided by
NSF grants AST-9980846 and AST-0206158.  ER is supported by a Discovery
Grant from NSERC of Canada.  NJE and MM were supported by NSF Grant
AST-1109116.

\bibliographystyle{apj_w_etal}

\end{document}